\documentclass{ar-1col}
\usepackage[numbers]{natbib}
\usepackage[letterpaper, portrait, margin=1.5in]{geometry}

\jname{Annu. Rev. Condens. Matter Phys.}
\jvol{AA}
\jyear{2017}
\doi{10.1146/((please add article doi))}

\begin{document}

\markboth{Coldea and Watson}{The electronic structure of FeSe}

\title{The key ingredients of the electronic structure of FeSe}

\author{Amalia I. Coldea$^1$ and Matthew D. Watson$^2$
	\affil{$^1$Clarendon Laboratory, Department of Physics,
	University of Oxford, Parks Road, Oxford OX1 3PU, United Kingdom; email: amalia.coldea@physics.ox.ac.uk}
\affil{$^2$Diamond Light Source, Harwell Campus, Didcot OX11 0DE, United Kingdom; email: matthew.watson@diamond.ac.uk}
	}

\begin{abstract}
FeSe is a fascinating superconducting material at the frontier of research
in condensed matter physics. Here we provide an overview
on the current understanding of the electronic structure of FeSe,
focusing in particular on its low energy electronic structure
as determined from angular resolved photoemission spectroscopy, quantum oscillations
and magnetotransport measurements of single crystal samples.
 We discuss the unique place of FeSe amongst iron-based superconductors,
 being a multi-band system exhibiting strong orbitally-dependent electronic correlations and
 unusually small Fermi surfaces, prone to different electronic instabilities.
 We pay particular attention to the evolution of the electronic structure
 which accompanies the tetragonal-orthorhombic structural distortion of the lattice around 90~K,
which stabilizes a unique nematic electronic state.
 Finally, we discuss how the multi-band multi-orbital nematic electronic structure has an impact on the understanding of the superconductivity, and show that the tunability of the nematic state with chemical and physical pressure will help to disentangle the role of different competing interactions relevant for enhancing superconductivity.
\end{abstract}

\maketitle

\tableofcontents

\section{INTRODUCTION}

FeSe is structurally the simplest of the iron-based superconductors, but is host to some of the richest physics.  Shortly after  high-$T_c$ superconductivity was found
 in iron-pnictide systems by Kamihara \textit{et. al.} \cite{Kamihara2008}, superconductivity in the Fe-chalcogenide compound FeSe below 8~K was discovered by Hsu \textit{et. al.} \cite{Hsu2008}.  All these materials
 share a common structural motif of a square lattice of Fe, bonded to pnictogen or chalcogen ions which sit alternately above and below the plane. The appearance of unconventional superconductivity in the Fe-based systems is commonly thought to arise from a spin fluctuation pairing mechanism, linked with the suppression of spin-density wave ordering observed in the parent compounds. However the unique properties of non-magnetic FeSe provides a challenging test-case for this view, and has generated intense detailed experimental and theoretical investigation of its electronic and magnetic properties.

The superconductivity of FeSe is remarkably tunable. Under applied pressure, $T_c$ reaches 36.7~K at 8.9 GPa \cite{Medvedev2009}. Furthermore, the FeSe layers
are held by weak van der Waals bonds that make them susceptible to mechanical exfoliation and chemical
intercalation. Ionic gating using a field-effect layer transistor in thin flakes of FeSe induces dramatic changes
in the carrier density and enhances superconductivity towards 43~K \cite{Lei2016}. Intercalation has the effect of separating the layers and also effectively dopes them with electrons, and a similarly enhanced $T_c$ can be achieved \cite{Sun2015,BurrardLucas2013}.

The recent renaissance in research on FeSe has been mainly driven by significant advances in materials development.
Firstly, the discovery of high temperature superconductivity in a monolayer FeSe  grown by molecular beam epitaxy on SrTiO$_3$ \cite{Wang2012monolayer,Wang2017review} provided a exciting new route
towards strongly enhanced superconductivity  over 65~K in an two-dimensional iron-based superconductor.
Secondly, it was found that the chemical vapor transport method  \cite{Bohmer2013,Chareev2013,Bohmer2016} could yield mm-sized plate-like single crystals free of impurity phases and close to stoichiometry, enabling detailed study of the intrinsic physics of bulk FeSe.

In this review we will discuss the electronic behaviour of FeSe in high quality single crystals in order
to reveal the key ingredients behind its unusual nematic state and superconducting properties. We will focus on the experimental insights provided by high resolution ARPES studies over a large
range of temperatures, combined with low temperature quantum oscillations and magnetotransport experiments, and discuss the interplay of order parameters in this intriguing system.


\subsection{Basic physical properties of FeSe}

\begin{figure}
	\centering
	\includegraphics[width=0.7\linewidth]{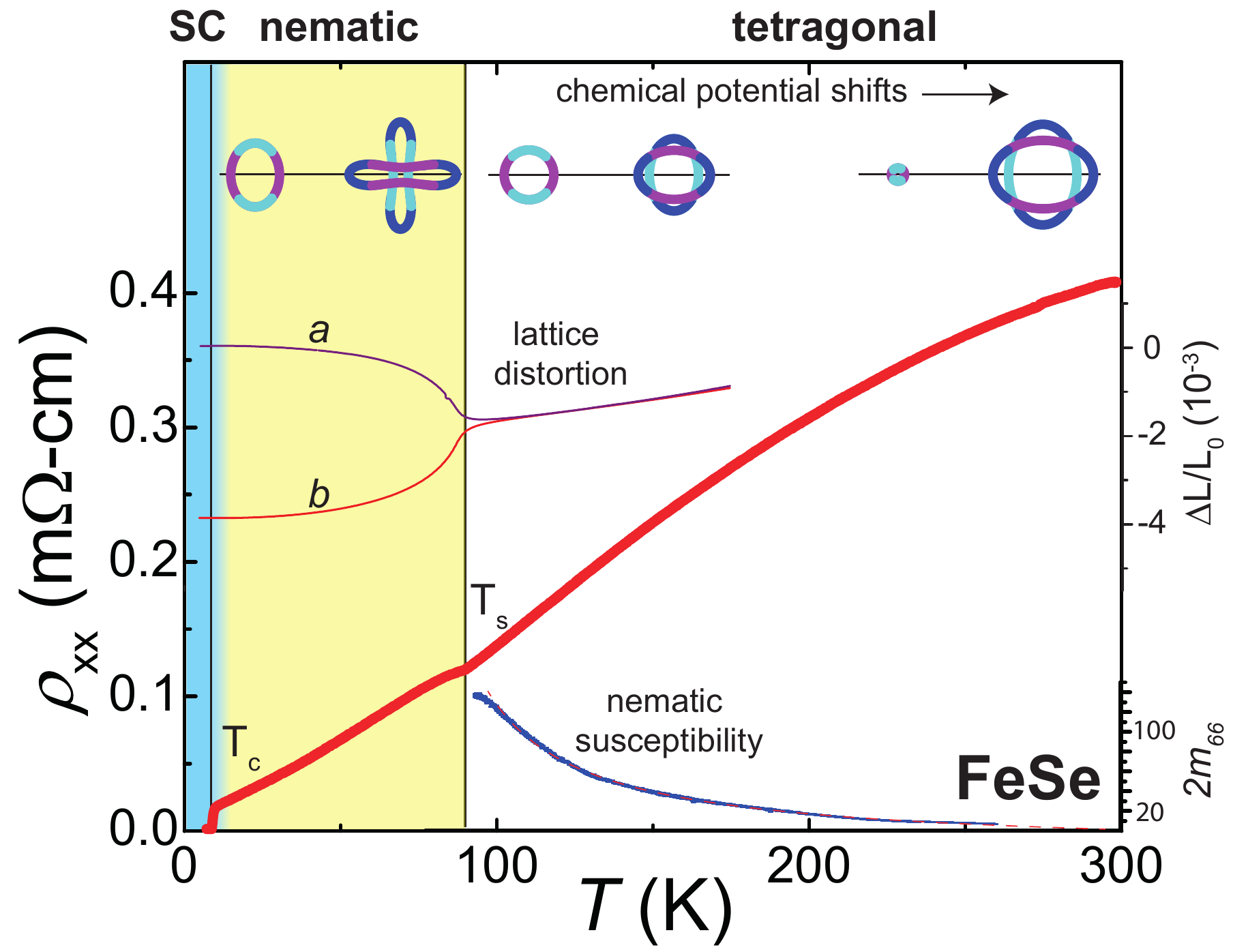}
	\caption{Temperature-dependent resistivity measurement of a single crystal of FeSe. Insets show thermal expansion data adapted from B\"ohmer \textit{et. al.} \cite{Bohmer2013} and nematic susceptibility data from Ref.~\cite{Watson2015a}. Schematic Fermi surfaces are based on ARPES measurements \cite{Watson2016,Rhodes2017_arxiv}.}
	\label{fig:r-t-schematic}
\end{figure}

{\bf Figure~\ref{fig:r-t-schematic}} show some of  the basic properties of FeSe at ambient pressure. At room temperature the resistivity starts to saturate around 0.4~m$\Omega$cm, similar to other Fe-based superconductors, which may indicate that the system approaches the Mott-Ioffe limit \cite{Gunnarsson2003},
when the mean-free path of electrons has become smaller than the in-plane lattice constant. In the low temperature
regime the resistivity display almost linear resistivity, before becoming superconducting at $T_c \sim 8.7$~K.

FeSe  exhibits a structural transition, from a tetragonal $P4/nmm$ to a weakly orthorhombic $Cmma$ unit cell at $T_s \sim$ 90~K  \cite{Mcqueen2009a,Khasanov2010}.
The magnitude of the structural distortion, as well as the preceding softening of the lattice, mimics other
 parent compounds of iron-based superconductors, like BaFe$_2$As$_2$ \cite{Bohmer2013}.  Other parent and underdoped compounds undergo a structural transition at high temperature followed closely by a magnetic transition, however FeSe is unusual in that no long-range magnetic order has been detected.

 The driving force of this symmetry-breaking transition is hotly debated \cite{Chubukov2015,Fernandes2014,Onari2016,Wang2015}. The transition at $T_s$ is unlikely to be
 driven by the lattice degrees of freedom due
 to the very small changes in in-plane lattice parameters \cite{Bohmer2013}, shown in {\bf Figure~\ref{fig:r-t-schematic}}.
 Instead, the breaking of fourfold rotational symmetry at $T_s$, is driven
 by the development of a {\it nematic} electronic state, characterised by strong in-plane anisotropy observed
 in  resistivity \cite{Tanatar2016} and quasiparticle interference \cite{Kasahara2014,Sprau2016_arxiv}.
 Important evidence for an electronically-driven transition
 is the divergence of the {\it nematic susceptibility} approaching $T_s$ \cite{Watson2015a,Hosoi2016},
 measured as the induced resistivity anisotropy in response to an external strain, shown in  {\bf Figure~\ref{fig:r-t-schematic}}.
 Thus we expect signatures of this electronic
 nematic state to manifest either in strongly anisotropy effects of its Fermi surface, often referred as a
 {\it Pomeranchuk instability} with $d$-wave symmetry \cite{Pomeranchuk1959,Chubukov2015} or charge nematic fluctuations \cite{Massat2016},
  and anisotropic scattering below $T_s$.

 The lack of long-range magnetic order, despite the presence of significant spin-fluctuations detected by neutron diffraction \cite{Wang2016}, is a significant puzzle. Amongst several possible explanations, it has been suggested that FeSe may be close to several competing magnetic instabilities \cite{Glasbrenner2015}, or may be a strongly fluctuating quantum paramagnet \cite{Wang2015}. On the other hand, it has been proposed that the structural transition may be non-magnetic in origin, instead being a manifestation of orbital ordering \cite{Onari2016,Chubukov2016}. The origin of the structural transition at $T_s$ is closely linked to the debate concerning the mechanism of superconductivity in FeSe, with wide implications on iron-based superconductivity in general. The nematic order is likely to significantly influence
the superconductivity, with recent evidence of orbitally-selective pairing and strongly anisotropic and twofold-symmetric superconducting gap structure \cite{Sprau2016_arxiv,Xu2016}. Thus, the understanding of these intriguing physical properties requires detailed knowledge about the experimental electronic structure in both the tetragonal and nematic phases.

\begin{marginnote}[]
	\entry{Nematic phase}{By analogy to the nematic phase of liquid crystals with broken rotational symmetry, the non-magnetic orthorhombic phase of FeSe is described as a nematic phase since fourfold rotational symmetry is broken while translation and time-reversal symmetries are preserved, although here the system has only two choices of orientation \cite{Fernandes2014}. }
\end{marginnote}

\section{ARPES STUDIES OF FESE}
\begin{figure}[t]
	\centering
	\includegraphics[width=0.95\linewidth]{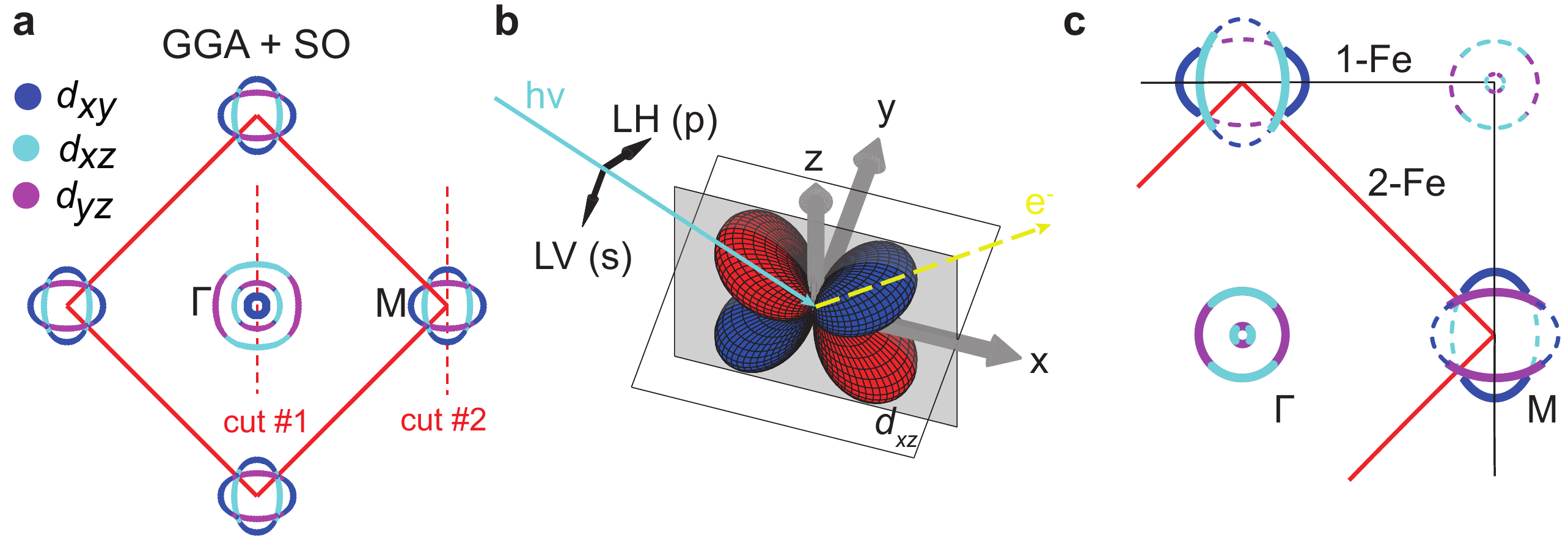}
	\caption[ARPES schematic]{a) DFT calculation of the Fermi surface of FeSe, from Ref.~\cite{Watson2015a}. b) Schematic diagram of the matrix elements considerations based on parities of atomic orbitals, in this case showing the $d_{xz}$ orbital, which would conventionally appear in LH polarisation only. c) The effect on the glide symmetry on the spectral weight as probed by ARPES, based on Brouet \textit{et. al.} \cite{Brouet2012}. Dashed lines indicate dispersions which are expected to have a suppressed spectral weight.}
	\label{fig:arpesschematic}
\end{figure}

\begin{textbox}[b]\section{What does ARPES actually measure in FeSe?}
	The $P4/nmm$ unit cell of tetragonal FeSe includes two Fe sites which are related by a glide symmetry. It has been argued that the essential physics can be captured in an effective 1-Fe unit cell with half the number of bands, constructed by {\it unfolding} the 2-Fe Brillouin zone \cite{YWang2015,Tomic2014}. This raises questions about what structure will be observed in ARPES measurements. Experimentally, ARPES measurements can detect the entire Fermi surface expected for the 2-Fe unit cell in FeSe at 100 K \cite{Watson2016,Fedorov2016}, and also in e.g. LiFeAs \cite{Borisenko2015}. However the underlying symmetry of the 1-Fe unit cell modulates the observations \cite{Brouet2012,Moreschini2014} such that the spectral intensity on some branches may be suppressed ({\bf Figure~\ref{fig:arpesschematic}c)}. On top of this, the intensity of features always depends strongly on the polarisation of the incident beam due to the different orbital characters and parities of the bands ({\bf Figure~\ref{fig:arpesschematic}b}). The effective $k_z$ depends on the incident photon energy, which also affects the relative intensity of features.
\end{textbox}

\begin{marginnote}[]
	\entry{ARPES}{Angle-resolved photoemission spectroscopy. This surface-sensitive technique involves the analysis of the energy and momentum of photoelectrons coming from a freshly cleaved sample surface.}
\end{marginnote}
Both thin films and high quality vapor-grown single crystals of FeSe are very suitable and popular systems for ARPES investigations, having large, flat and non-polar (001) surfaces. Here, we discuss mainly ARPES measurements of single crystals of FeSe with a specific focus on understanding of the nematic ordering,
rather the case of the high $T_c$ monolayer FeSe on different substrates reviewed elsewhere \cite{Pustovit2016,Liu2015a,Wang2017review}.
The many other ARPES studies on related iron selenides, including FeTe$_{1-x}$Se$_x$, alkali-metal dosed FeSe, electron-doped (K,Rb)$_x$Fe$_{2-y}$Se$_2$, and (Li$_{0.8}$Fe$_{0.2}$)OHFeSe are reviewed elsewhere \cite{Pustovit2016,Liu2015a,Wang2017review,Huang2017_arxiv}, and general overviews on ARPES studies of iron-based superconductors
may be found in Refs.~\cite{Richard2011,Kordyuk2012,Richard2015,vanRoekeghem2016}.

\subsection{The hole pockets}

\begin{figure}[t]
	\centering
	\includegraphics[width=0.9\linewidth]{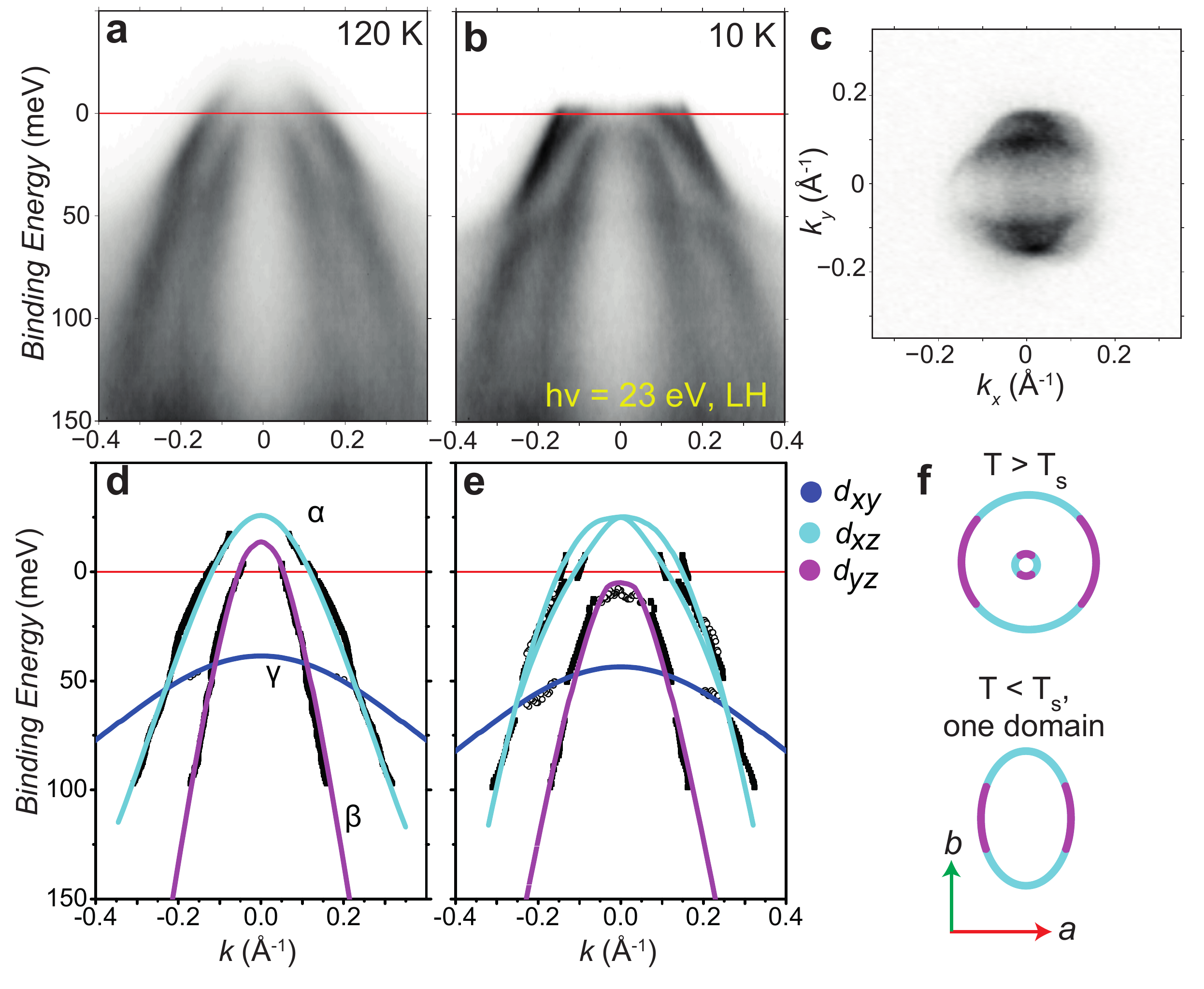}
	\caption[Hole pockets]{a) High and b) low temperature spectra at the Z point (23 eV) , and c) low temperature Fermi surface map at Z. d,e) Extracted band positions, overlaid with guides to the eye indicating the band dispersions. f) Schematic diagram of the hole pockets in the tetragonal and orthorhombic phases. The anisotropic distortion
of the Fermi surface is a signature of a nematic electronic phase.
Figures adapted from Ref.~\cite{Watson2015a}.}
	\label{fig:holepockets}
\end{figure}

According to DFT calculations ({\bf Figure~\ref{fig:arpesschematic}a}), FeSe should exhibit three quasi-2D hole pockets at the zone centre, which occupy a sizeable fraction of the Brillouin zone, but the experimental picture varies substantially. The observed dispersions are significantly renormalised compared to the DFT dispersions, and shifted down such that the pockets are much smaller \cite{Watson2015a}. In particular the $d_{xy}$ band is much flatter than expected, with a renormalisation factor of 8 \cite{Maletz2014,Watson2015a}, and is found at $\sim$50 meV below the Fermi level. The hole pockets are quasi-2D, with rather small $k_F$ values at the $\Gamma$ point, becoming larger at the Z point. When measured in the tetragonal phase above 90~K ({\bf Figure~\ref{fig:holepockets}a}), the Fermi surface of FeSe consists of an outer quasi-2D hole pocket and a small 3D inner hole pocket which just crosses the Fermi level around the Z point \cite{Watson2015a}. These Fermi surface are essentially circular, and come from the $d_{xz/yz}$ bands which are split at the $\Gamma$/Z point by spin-orbit coupling only \cite{Watson2015a,Fernandes2014b}, with a band separation $\Delta_{SO}\sim$~20~meV \cite{Watson2015a,Watson2017c_arxiv,Borisenko2015}. In the nematic phase, the Fermi surface distorts substantially into an elliptical shape, and in addition the inner hole pocket moves completely below the Fermi level. Thus the low temperature Fermi surface of FeSe around the zone centre consists of a single elliptical quasi-2D band ({\bf Figure~\ref{fig:holepockets}f}). Since samples will naturally form twin domains below 90 K, two superposed ellipses are observed in most ARPES measurements ({\bf Figure~\ref{fig:holepockets}e}), though a single ellipse may be observed in detwinned measurements \cite{Shimojima2014,Suzuki2015}.

\subsection{The electron pockets}

\begin{figure}
	\centering
	\includegraphics[width=1\linewidth]{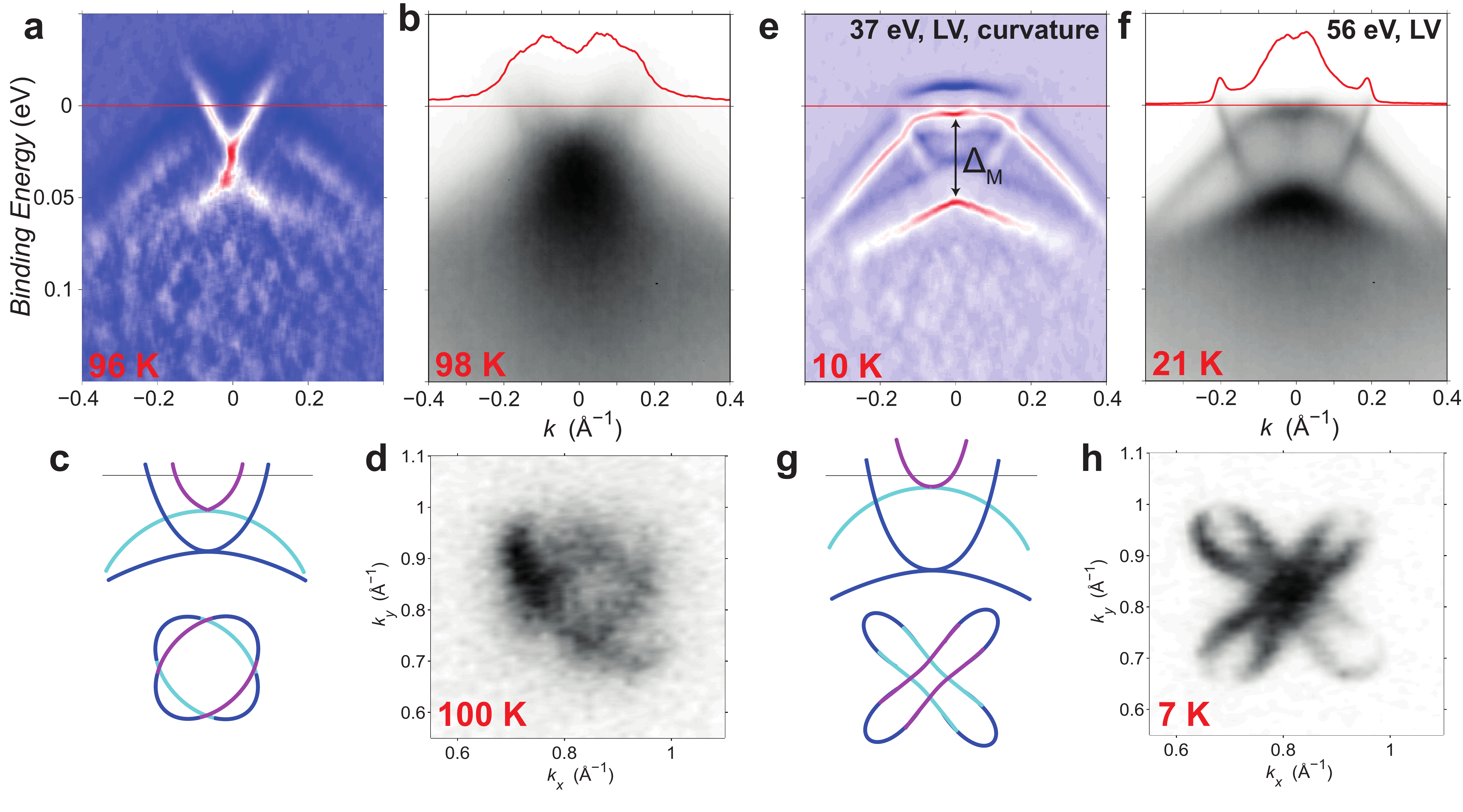}
	\caption[Electron pockets]{a) ARPES measurements in the high-symmetry direction through the M point, above $T_s$, showing curvature data at 37 eV. b) ARPES measurements at 56 eV in the same orientation. The MDC at the Fermi level in the 56 eV data consists of four peaks. c) Schematic electronic structure in the tetragonal phase. d) Fermi surface map at 100 K, above $T_s$. e-g) Equivalent measurements at low temperatures.}
	\label{fig:electronpockets}
\end{figure}

The electron pockets around the M point are also quasi-2D and much smaller than expected in the DFT calculation,  also showing renormalisation by a factor of $\sim$ 4 \cite{Watson2015a} for the $d_{yz}$ band in the tetragonal phase. {\bf Figure~\ref{fig:electronpockets}c} shows that at low temperature, the high-symmetry cut through the M point contain four bands, with two features in the Energy Dispersion Curve (EDC) at the M point separated by a famously large energy scale of $\Delta_M\sim$50 meV ($\sim$600~K).

The understanding of the ARPES spectra at the M point has been a contentious issue, and some historical context is useful. Pioneering work on multilayer thin film samples of FeSe \cite{Tan2013} obtained spectra qualitatively similar to {\bf Figure~\ref{fig:electronpockets}c}. Due to the analogy with similar features seen on an equally large energy scale in magnetic parent compounds NaFeAs and BaFe$_2$As$_2$ around the M point, the 50 meV energy scale was first associated with spin density wave order in thin films  \cite{Tan2013}, but later on in single crystals was linked to a nematic ordering \cite{Nakayama2014}, due to the lack of long-range magnetic order in FeSe. It was suggested that in the nematic phase the $d_{xz/yz}$ band degeneracy at the M point is dramatically lost, and the 50~meV energy scale is the magnitude of the $d_{yz}-d_{xz}$ orbital ordering (at M), with the onset of band shifts occurring at $T_s \sim 90$~K \cite{Shimojima2014,Watson2015a}.

 Recent improvements in the data quality, particularly at temperatures above 90~K, clearly point towards a different interpretation of the band dispersions at the M point.  The crucial evidence is the observation of distinct $d_{xy}$ dispersions in data obtained above 90~K \cite{Watson2016,Fedorov2016,Fanfarillo2016,Zhang2016}. In {\bf Figure~\ref{fig:electronpockets}a} the curvature analysis reveals clear evidence for the lower branch of the $d_{xy}$ dispersion at M in addition to the $d_{xz},d_{yz}$ dispersions, while the branch forming the outer $d_{xy}$ section of the Fermi surface is clearly seen in data obtained at 56 eV in {\bf Figure~\ref{fig:electronpockets}b} and {\bf d}. The fact that $d_{xy}$ dispersions also contribute brightly to the observed spectra rules out the assignment of the 50 meV splitting to only the $d_{xz}$ and $d_{yz}$ bands. Instead it is the separation of $d_{xy}$ and $d_{xz/yz}$ bands at the M point. This band separation does appear to increase in magnitude below 90~K and is therefore linked with nematic order, but is not intrinsically a measure of any orbital symmetry-breaking.
\begin{figure}
	\centering
	\includegraphics[width=0.95\linewidth]{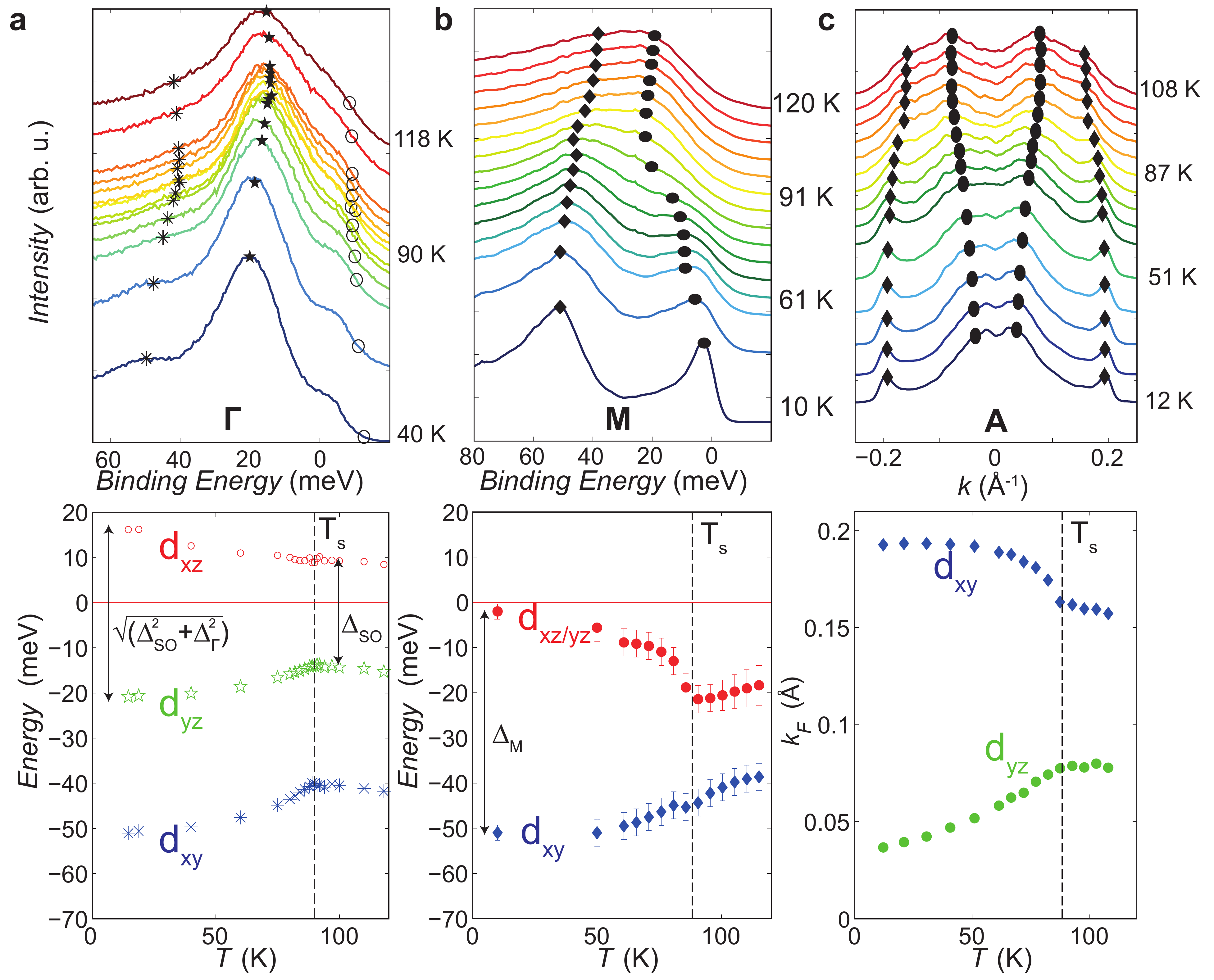}
	\caption{a) Temperature-dependence of EDCs at the $\Gamma$ point, lower panel showing extracted band positions from a fitting analysis. b) Similar analysis at the M point. c) Analysis of MDCs at the A point. Figure adapted from Refs.~\cite{Watson2016} and \cite{Watson2017c_arxiv}.}
	\label{fig:reviewfigarpestdepdata}
\end{figure}
\subsection{Temperature-dependence of the position of the bands}

\begin{figure}
	\centering
	\includegraphics[width=0.95\linewidth]{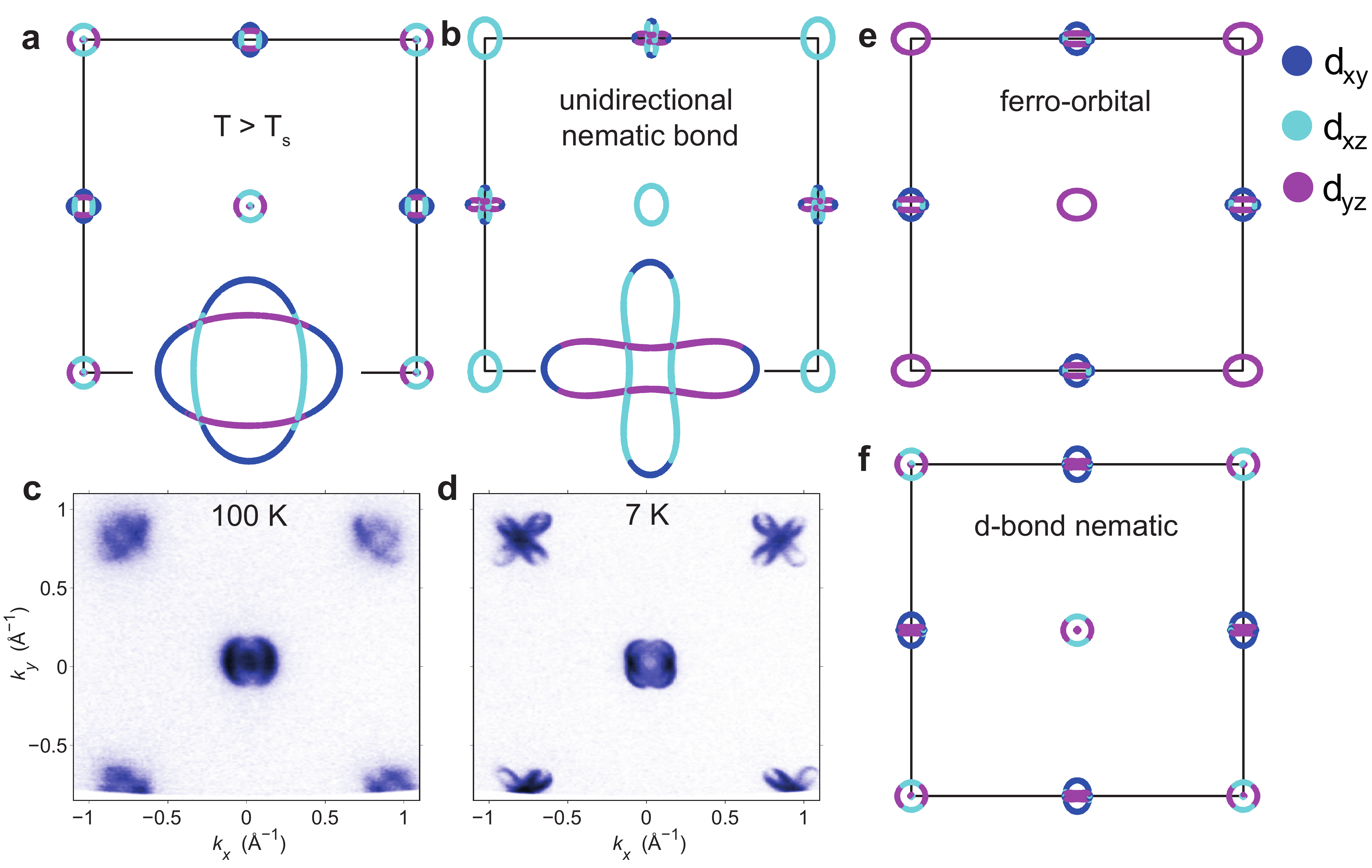}
		\caption[Electron pockets]{a) Tight binding simulations of the Fermi surface in the tetragonal phase, closely based on the experimental dispersions. b) Calculated Fermi surface including the unidirectional nematic bond ordering term. c,d) Experimental Fermi surfaces at high and low temperatures. e,f) Other proposed orbital ordering scenarios, the ferro-orbital ordering and $d$-wave bond nematic orders. Figures adapted from Ref.~\cite{Watson2016}. }
	\label{fig:orderparameters}
\end{figure}

We now focus on the details of the band shifts in the nematic phase, with the aim of determining the magnitude and momentum-dependence of the nematic order parameter. We start with the hole pockets at high temperatures in the tetragonal phase, where at the $\Gamma$ point the $d_{xz}$/$d_{yz}$ bands are separated by spin-orbit coupling only \cite{Fernandes2014b}, with the magnitude of this separation being $\sim$20 meV \cite{Watson2015a,Borisenko2015,Watson2017c_arxiv}. While Zhang \textit{et. al.} found no significant temperature-dependence at the $\Gamma$ point, most studies agree that there is an increased separation of the $d_{xz}/d_{yz}$ hole bands in the nematic phase \cite{Watson2015a,Fedorov2016,Watson2016}, which was studied in detail in Ref.~\cite{Watson2017c_arxiv}, reproduced in {\bf Figure~\ref{fig:reviewfigarpestdepdata}a}. As the spin-orbit term is expected to be temperature-independent, this increase in splitting is the signature of a symmetry-breaking orbital order parameter. It is important to note that the band separation due to spin-orbit coupling will add in quadrature, not linearly, with the extra splitting associated with the nematic order at the $\Gamma$ point \cite{Fedorov2016,Fernandes2014,Watson2017c_arxiv}. From measurements of twinned samples, it is impossible to determine whether it is the $d_{xz}$ or $d_{yz}$ orbital which is raised/lowered in energy at the $\Gamma$ point. This information is obtained by {\it detwinning} the samples using a mechanical strain \cite{Shimojima2014,Suzuki2015}, from which it may be deduced that the $d_{xz}$ orbital is raised in energy while the $d_{yz}$ orbital is lowered in energy ({\bf Figure~\ref{fig:orderparameters}a}). Extracting the position of the $d_{xz}$ band in the unoccupied states is a difficult task, but in Ref.~\cite{Watson2017c_arxiv} a value of 37.5 meV was estimated, implying that the symmetry-breaking component $\Delta_\Gamma \approx$ 29 meV after the spin-orbit coupling is accounted for. The downward shift of the $d_{yz}$ band brings the small 3D hole pocket seen at the Z point completely below the Fermi level ({\bf Figure~\ref{fig:holepockets}b}). Therefore the low temperature Fermi surface at the zone centre consists of a single elliptical quasi-2D hole pocket, with the longer direction of the hole pocket oriented along the shorter $b$ axis of the orthorhombic structure ({\bf Figure~\ref{fig:holepockets}f}).

\begin{marginnote}[]
	\entry{EDC and MDC}{Energy Distribution Curves and Momentum Distribution Curves, slices of the intensity maps over binding energy and momentum $k$ generated by ARPES.}
\end{marginnote}

{\bf Figure~\ref{fig:reviewfigarpestdepdata}b} shows the EDCs at the M point as a function of temperature.
In Ref.~\cite{Watson2016} it was shown based on fitting analysis that above 90~K there are actually two features in the spectra, which are separated by $\sim$ 20 meV.
These must be associated with the separate $d_{xz/yz}$ and $d_{xy}$ bands which are expected at M ({\bf Figure~\ref{fig:electronpockets}c}). Below 90~K, in the orthorhombic phase the band degeneracies at the M point are no longer protected by fourfold symmetry, thus it would be natural to expect that the $d_{xz}$ and $d_{yz}$ dispersions should separate, which would give extra features in the EDC at the M point in a twinned sample. Such a separation was also observed in Ref.~\cite{Fedorov2016}, on the order of 10 meV. However most data sets at the M point show only two features even at the lowest temperatures \cite{Shimojima2014,Watson2015a,Watson2016}. If there are only two features, this would imply an unexpected situation: the $d_{xz/yz}$ and $d_{xy}$ bands have an increased separation below 90~K, reaching $\sim$50 meV by low temperatures, but no $d_{xz/yz}$ splitting is resolved. It is helpful to additionally analyse the MDCs as a function of temperature. This analysis is best performed with an incident photon energy of 56 eV, ({\bf Figure~\ref{fig:reviewfigarpestdepdata}c}), where the outer branch of the electron pockets is easily distinguished at temperatures above 90~K. By following the features in the MDC by fitting with Lorentzian peak profiles, it can be easily deduced that the $d_{yz}$ $k_F$ shrinks substantially, while the outer $d_{xy}$ section increases below 90~K, so the pocket evolves from the two crossed ellipses seen at high temperature into the crossed peanut-shaped (or {\it bow-tie shaped} \cite{Sprau2016_arxiv}) bands at low temperature.


\subsection{The orbital order in FeSe}

We now consider the implications of these observations on the determination of the form of the orbital order parameter. The simplest order parameter would be a momentum-independent orbital polarization, or ferro-orbital ordering, $\frac{\Delta}{2}(n_{yz}-n_{xz})$, shown in {\bf Figure~\ref{fig:orderparameters}e}. Early on, based primarily on the M point data several groups proposed a 50~meV ferro-orbital ordering \cite{Shimojima2014,Nakayama2014,Watson2015a,Watson2015b,Kreisel2015}. However, this interpretation has been revised, as this effect is of much smaller magnitude at the $\Gamma$ point \cite{Watson2015a}. Moreover, ARPES studies under strain have established that the $d_{yz}$ band moves in opposite directions at the $\Gamma$ and at the M points \cite{Suzuki2015}, which clearly points towards a momentum-dependent order parameter. If ferro-orbital ordering is excluded, or at least it is known to be not the primary order parameter of the transition, one must start to consider bond-type orderings \cite{Jiang2016}, such as a $d$-wave bond nematic order \cite{Jiang2016,Yi2015,Zhang2015}, plotted in {\bf Figure~
\ref{fig:orderparameters}f}. However since a pure $d$-wave symmetry order parameter would give no extra splitting at the $\Gamma$ point \cite{Jiang2016}, this is excluded. A composite order parameter involving both ferro-orbital and $d$-wave bond nematic order parameter has been recently discussed \cite{Sprau2016_arxiv,Scherer2017}, which could perhaps account for the $\Gamma$ point data with some fine-tuning, but would give a $d_{xz/yz}$ splitting at the M point.
\begin{marginnote}[]
	\entry{Orbital order}{Here we use this term to refer to a generic non magnetic symmetry-breaking order parameter, which may be written in terms of on-site orbital energies or orbital hoppings, and manifests in electronic band shifts. It is a different usage to the well-known orbital ordering effect associated with Jahn-Teller physics.}
\end{marginnote}

Another candidate is the {\it unidirectional nematic bond order},  introduced in Ref.~\cite{Watson2016}, which provides a simple and accurate description of the band shifts and splitting in the nematic phase. This order parameter, written as ${\Delta_u}(n'_{yz}-n'_{xz})\cos(k_x)$,  does not affect the on-site orbital energies, but instead breaks fourfold symmetry in the inter-site hopping terms (i.e. $n'$ indicates a hopping term) \cite{Watson2016}. It has the effect of giving an extra splitting of $d_{xz}$-$d_{yz}$ bands at the $\Gamma$ point,  giving a total separation of $\sqrt{\Delta_{SO}^2+(2\Delta_u)^2}$, but shifting the $d_{yz}$ and $d_{xz}$ dispersions up together by $\Delta_u$ at the M point, without breaking the degeneracy there. The experimental data at the $\Gamma$ \cite{Watson2017c_arxiv} and M \cite{Watson2016} points would indicate values of $\Delta_{u,\Gamma} \approx$14.5 and $\Delta_{u,M} \approx$ 20 meV, consistent within the experimental uncertainty. The Fermi surface and band dispersions with and without this order parameter ({\bf Figures~\ref{fig:orderparameters}a,b}), which show an good correspondence with the experimental data in both phases. However at the present time a microscopic motivation for this order parameter is lacking. Moreover high resolution detwinned ARPES measurements will help to eliminate uncertainty over the role of different domains, particularly at the M point. Finally we note that with all these scenarios, a small contribution from other symmetry-allowed order parameters is to be expected, for example a small ferro-orbital ordering contribution cannot be ruled out in addition to the unidirectional nematic bond order.

\begin{marginnote}[]
	\entry{Pomeranchuk instability}{is a term
		use to reflect the development
		of the nematic behaviour due to the
		distortion of the Fermi surface of a metal
		in the presence of
		strong electronic correlations.
		In FeSe,
		the type of Fermi surface distortion
		is often referred as a
		$d$-wave Pomeranchuk instability.}
\end{marginnote}

\subsection{FeSe as a strongly correlated system}

So far we have focused on the $\sim$20 meV band shifts associated with nematic ordering, but a complete picture of the physics of the system includes an understanding of the few-eV energy scale. FeSe is a strongly correlated material, exhibiting an enhancement of the quasiparticle effective masses in quantum oscillations \cite{Shimojima2014,Watson2015a,Watson2015b} and the Sommerfeld coefficient \cite{Bohmer2014}. However the electron-electron interactions which are primarily responsible for these effects around the Fermi level also manifest on energy scales comparable with the Fe $3d$ bandwidth, which can be probed by measuring photoemission spectra across a wide range of binding energies.

\begin{figure}
	\centering
	\includegraphics[width=0.95\linewidth]{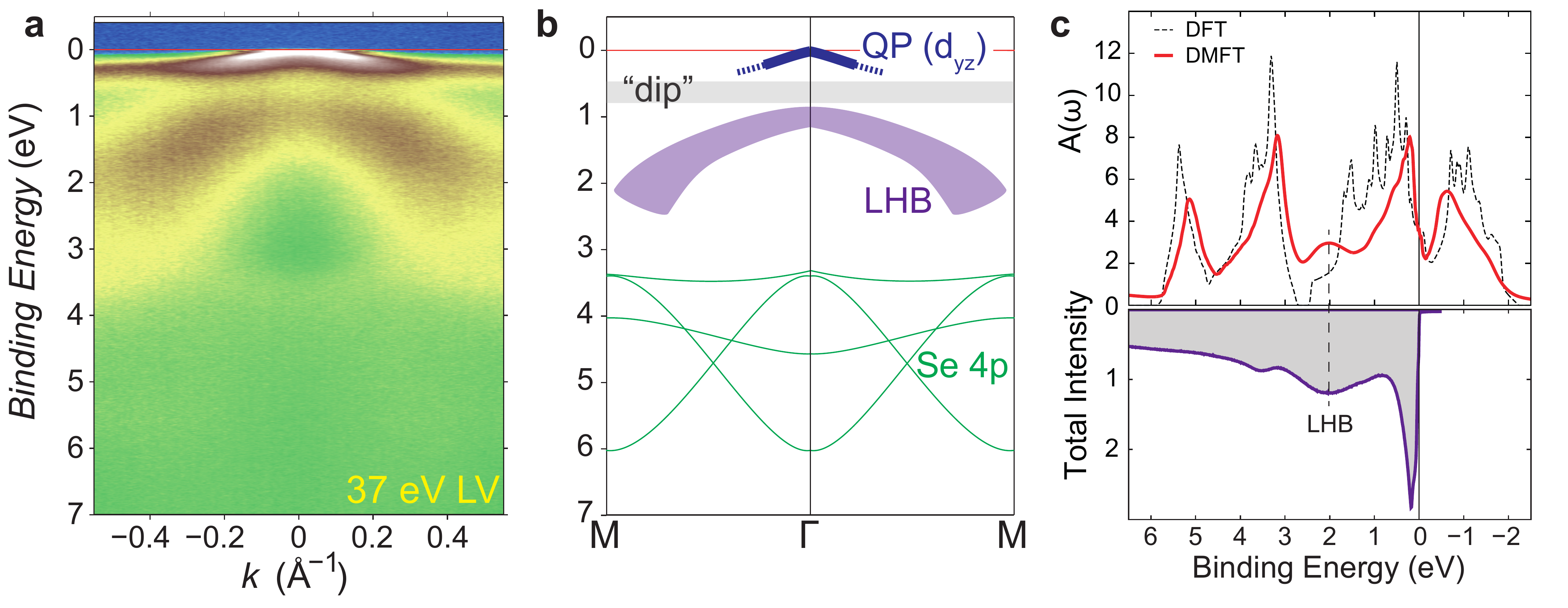}
	\caption[High Energy spectral function of FeSe]{a,b) ARPES spectra measured through the $\Gamma$ point, plotted on a wide energy scale. b) Schematic understanding of features seen in panel a). QP and LHB refer to quasiparticle dispersions and the Lower Hubbard Band. Hints of the Se $4p$ dispersions are also seen. c) Comparison of total ARPES intensity with DFT and DFT+DMFT calculations. Figures adapted from Ref.~\cite{Watson2017a}}
	\label{fig:reviewfigarpeshighenergy}
\end{figure}

In this context, it is worth emphasizing that the quantity probed by ARPES is the one-particle spectral function, mulitplied by the relevant photoemission matrix element and the Fermi function. The spectral function contains very rich information about many-body effects in the system. Around the Fermi level, the ARPES spectra show sharp quasiparticle bands, which have primarily $d_{xz}$,$d_{yz}$ and $d_{xy}$ orbital character. {\bf Figure~\ref{fig:reviewfigarpeshighenergy}a} shows an ARPES measurement obtained in LV polarisation at the $\Gamma$ point, which highlights the $d_{yz}$ spectral weight. It can be seen that the quasiparticle bands give way to a featureless region giving a dip in the total spectral intensity around $\sim$0.5 eV binding energy, after which a dispersive but much broader feature is seen, around 1-2.5 eV binding energy. With a width of $\sim$1 eV, this can hardly be described as a quasiparticle band, but rather should be considered as incoherent spectral weight, or a {\it Hubbard band} \cite{Watson2017a,Evtushinsky2016_arxiv}. At $\sim$3-6 eV binding energies, hints of the Se $4p$ band dispersions are seen in some measurement geometries \cite{Watson2017a,Evtushinsky2016_arxiv}, very close to their expected locations according to DFT calculations.

As shown in {\bf Figure~\ref{fig:reviewfigarpeshighenergy}c}, this structure in the Fe $3d$ bandwidth is captured well within the DFT+DMFT technique, which accounts for the large on-site Coulomb repulsion term $U$=4 eV and the Hund's rule interaction $J_H$=0.8 eV which are not adequately treated within DFT alone. The calculated total spectral function exhibits a similar peak-dip-hump structure to the experimental data. Not every detail can be accounted for in the DFT+DMFT calculations, and in particular both DFT and DFT+DMFT predict much larger sizes of the Fermi surfaces than are seen experimentally. The general picture is that FeSe is a strongly correlated material, in which the on-site $U$ and $J_H$ play an important role in shaping the overall structure of the $3d$ bandwidth, but other effects such as the interatomic Coulomb interactions $V$ \cite{Jiang2016,Scherer2017} may also be relevant for understanding finer details of the electronic structure.
\begin{marginnote}[]
	\entry{DFT+DMFT}{Density Functional Theory combined with Dynamical Mean Field Theory, an \textit{ab-initio} approach to strongly correlated materials.}
\end{marginnote}

\section{QUANTUM OSCILLATIONS IN FESE}

\begin{figure}[h]
	\centering
	\includegraphics[trim={1cm 10cm 1cm 1cm},width=1\linewidth]{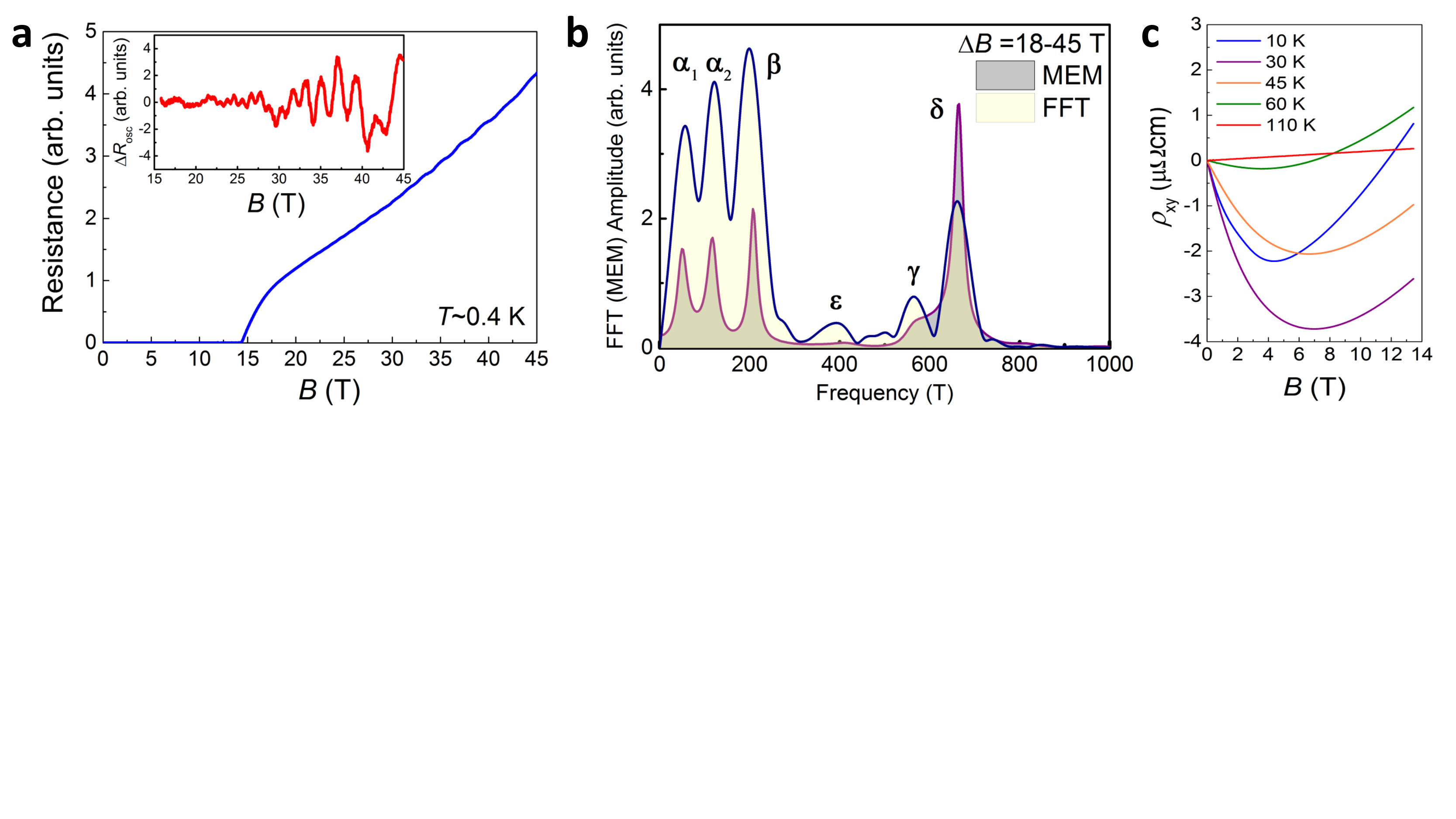}
	\caption{a) Magnetotransport data in a single crystal of FeSe at low temperatures and field up to 45~T.
		The inset shows the extracted quantum oscillatory signal. b) Fast Fourier transform (FFT) obtained using a Hanning window, together
		with the maximum entropy method (MEM) signal, allow to identify the presence of different cross-sectional areas
		on the Fermi surface. c) Hall effect measurement in low magnetic fields up to 13~T  show non-linear behavior as a function of temperature.}
	\label{fig:QOs}
\end{figure}

The observation of quantum oscillations  in single crystals of FeSe grown using chemical vapor transport
 have been reported by different groups \cite{Terashima2014,Audouard2014,Watson2015a,Watson2015b}, using magnetotransport and tunnel diode oscillator techniques.
{\bf Figure~\ref{fig:QOs}} shows the transverse magnetoresistivity as a function of magnetic field up to 45~T at 0.4~K for a single crystal of FeSe with
a residual resistivity ratio larger than 25.
Shubnikov-de Haas oscillations are detected in the normal state above 18~T on top of the strongly magnetoresistive background
and a complex oscillatory signal is revealed after the background subtraction in the inset of {\bf Figure~\ref{fig:QOs}a}.
The fast Fourier transform (FFT) spectra show
a large number of closely spaced peaks corresponding to quantum oscillation frequencies below 700~T ({\bf Figure~\ref{fig:QOs}b}).
These frequencies are significantly smaller than those seen in
other highly crystalline iron-based superconductors
such as LiFeAs, LiFeP and BaFe$_2$(As$_{1-x}$P$_x$)$_2$, suggesting that the low temperature Fermi surface of
FeSe has small pockets, a factor five smaller that the largest frequency predicted by DFT calculations of FeSe \cite{Watson2015a},
and occupying less than 2.3\% of the first Brillouin zone \cite{Terashima2014}.

\begin{figure}[t]
	\centering
	\includegraphics[width=1\linewidth]{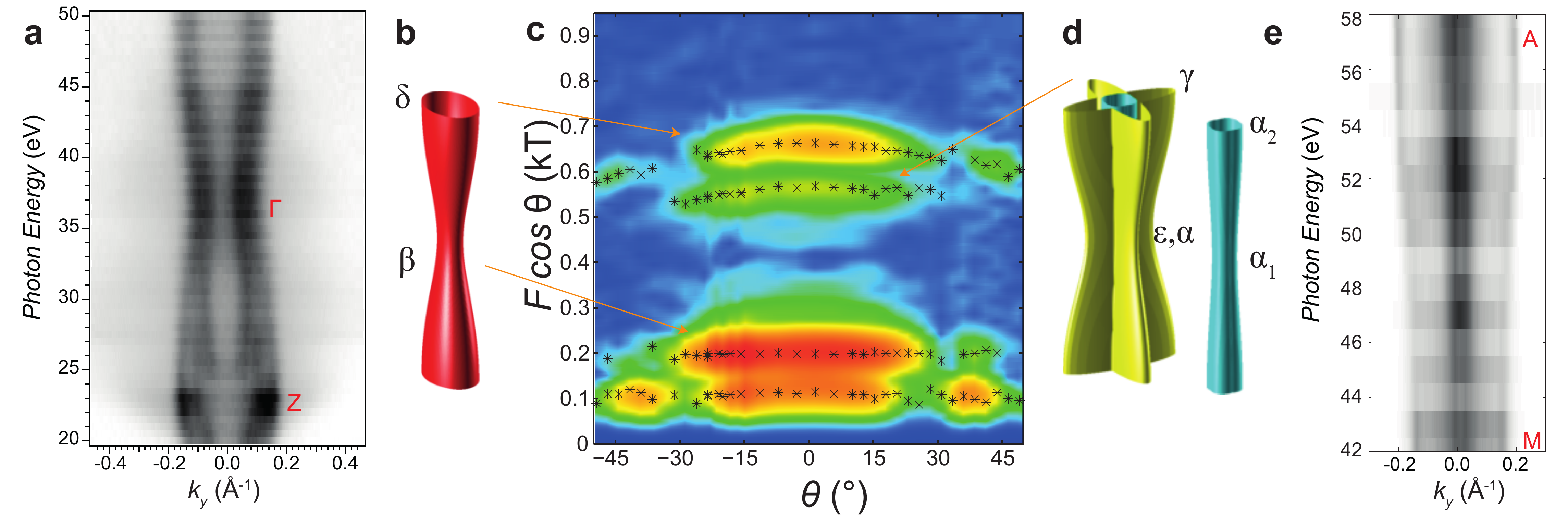}
	\caption{Three-dimensional representation of the low temperature quasi-2D Fermi surfaces of FeSe, from photon-energy dependent ARPES measurements for the hole band in a) and electron bands in e) . The schematic representation of the Fermi surface in (b,d) to illustrate the position of the
		maximum and minima extremal orbits determined in quantum oscillations.  c) Angular dependence of the quantum oscillation frequencies;
		intensity map represents the amplitude of oscillations.}
	\label{fig:kz-dispersion}
\end{figure}

\begin{textbox}[b]\section{What can be learned from quantum oscillations?}
	Quantum oscillations is a well-established and powerful technique for the
	experimental characterisation of the Fermi surface at low temperatures. Due to the Landau quantisation of electronic states in an applied magnetic field $B$, oscillations of various physical properties  periodic in $1/B$  are observed ({\bf Figure~\ref{fig:QOs}a}). The frequency of oscillation relates to extremal areas of the Fermi surface,  and the temperature-dependence of the amplitude of oscillations reveals the orbitally-averaged 	quasiparticle masses.
	
	The first major constraint for the observation of quantum oscillations is that the cyclotron energy which separates Landau levels needs to be larger than
	the broadening of the levels $\hbar/\tau$ due to scattering Thus only very high quality single crystals with long mean free paths show quantum oscillations.
Secondly, the amplitude of the quantum oscillations is significantly damped for 	heavier quasiparticle masses as a result of the smearing of the Landau levels by the Fermi-Dirac distribution.
Lastly, in order to observe quantum oscillations in a superconducting system, the superconductivity needs to be suppressed by a magnetic field which exceeds the upper critical field, since the gapping of quasiparticle states very strongly suppresses any oscillatory signal.
	
	While quantum oscillations give values of the extremal cross sectional areas with a greater precision than ARPES, the assignment of quantum oscillation frequencies to different pockets in $k$-space can only be done by comparison with theoretical modelling or other techniques, and can be a complex problem in the presence of several Fermi surface pockets. However the shape of the Fermi surface can be determined from the angular dependence of these frequencies  \cite{Terashima2014,Watson2015a}. For quasi-2D pockets it is instructive to plot $F\cos{\theta}$ as a function of $\theta$, the angle between the applied field and the sample normal ({\bf Figure~\ref{fig:kz-dispersion}c}). On such a plot, a flat line would correspond to a perfect cylinder, while minimal (maximal) extremal areas of a warped quasi-2D band would have upward (downward) curvature.
\end{textbox}

In ultra-high magnetic fields, magnetotransport and Hall effect data
provided an interesting insight into the origin of the largest-amplitude $\beta$,$\gamma$ and $\delta$ peaks \cite{Watson2015b}.
By considering the relative amplitudes of the quantum oscillations of the
$\rho_{xx} $ and $\rho_{xy} $ components, together with the positive sign of the high-field Hall signal at very low temperatures, it was revealed that the single hole band is associated with the $\beta$ and $\delta$ extremal areas in {\bf Figure~\ref{fig:QOs}b} and
{\bf Figure~\ref{fig:kz-dispersion}}. These bands also have similar effective masses of around 4~$m_e$
~\cite{Terashima2014,Audouard2014,Watson2015a}.
This assignment is in good agreement with
the areas estimated from ARPES, with $k_F$ values varying from 0.1-0.15 \AA$^{-1}$ around the hole pocket at Z \cite{Watson2015a}.
Thus the hole band gives small carrier density of 3.58$\times 10^{20}$ cm$^{-3}$
and contributes 3.1(4) mJ/mol K$^2$ to the electronic specific heat.

\begin{figure}
	\centering
	\includegraphics[width=1\linewidth]{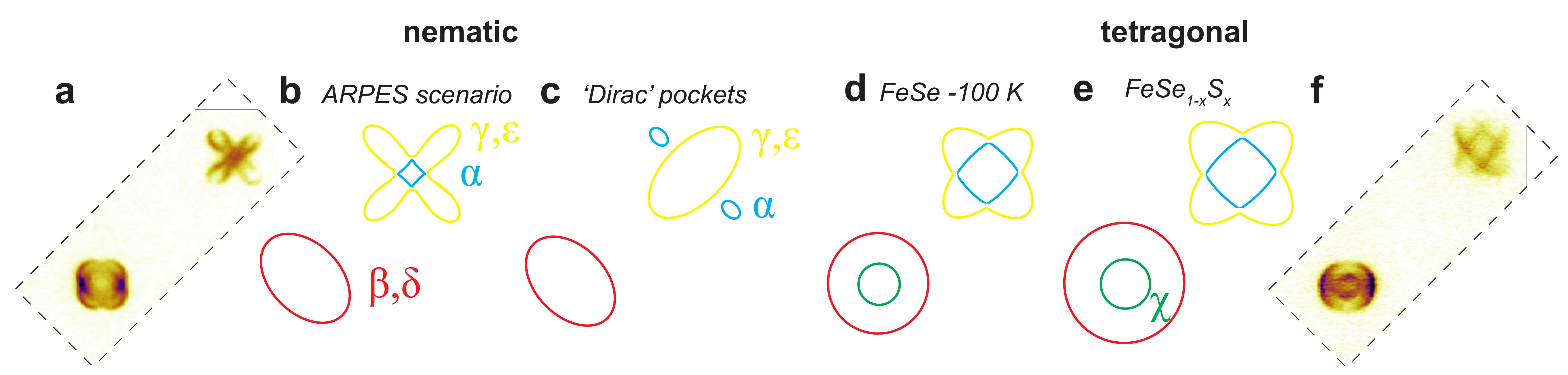}
	\caption{a) Low temperature Fermi surface map of FeSe \cite{Watson2016}. b) Possible orbits on the Fermi surface responsible for the observed quantum oscillation frequencies in FeSe and FeSe$_{1-x}$S$_{x}$ \cite{Watson2015a,Coldea2016}. c) An alternative scenario of the electron pockets which could give tiny electron pockets with a Dirac-like dispersion. d) Schematic Fermi surface of FeSe at high temperatures in the tetragonal phase and e) for a FeSe$_{1-x}$S$_{x}$ sample where nematicity is fully suppressed. f) Low temperature tetragonal Fermi surface of FeSe$_{0.82}$S$_{0.18}$ \cite{Reiss2017}.}
	\label{fig:FS_evolution_nematicity}
\end{figure}

The $\gamma$ frequency, the maximum of a quasi-2D pocket associated with the outer electron orbit, has
a notably heavier effective mass of $\sim 7(1)$ $m_e$ \cite{Watson2015a,Terashima2014}. The minimum orbit of this pocket has been assigned either to $\alpha_1$ \cite{Terashima2014}, suggesting that only one electron pocket is present,
or alternatively associated to the weak $\epsilon$ peak \cite{Coldea2016}, which
also account for the presence of an additional very small electron pocket ($\alpha_1$ and $\alpha_2$)
( {\bf Figure \ref{fig:kz-dispersion}}). The presence of a large electron pocket, having an almost compensated carrier density to the hole bands,
as well as that of a second tiny electron pocket was also determined from magnetotransport studies in FeSe \cite{Watson2015b}.

We now consider the different possible scenarios for the origin of the quantum oscillations associated with the electron pockets. In general, the orbits that are detected by quantum oscillations for the electron bands in non-magnetic iron-based superconductors originate from the inner and outer electron bands, separated by a sizable
spin-orbit hybridisation ({\bf Figure~\ref{fig:FS_evolution_nematicity}d,e}), as seen in LaFePO, LiFeAs and LiFeP \cite{Coldea2008,Putzke2012}. However, in FeSe the impact of nematic ordering, and the sensitivity of the Fermi surface to very small changes of the chemical potential, complicates the interpretation of the data. From the ARPES perspective, $\gamma$ frequency could be assigned to the four-leaf clover shaped orbit at the A point, shown in {\bf Figure~\ref{fig:FS_evolution_nematicity}b}.
Alternatively, due to the strongly distorted electron Fermi surface, this frequency could originate from other trajectories
either caused by breakdown orbits or other effects induced by the strong magnetic fields and at very low temperatures.
Small changes in the band positions relative to the chemical potential (1--2~meV)
 could push the inner electron bands above the Fermi level at the M point ({\bf Figure~\ref{fig:reviewfigarpestdepdata}}),
and lead to single elliptical orbit and two tiny, {\it Dirac-like}, electron pockets ({\bf Figure~\ref{fig:FS_evolution_nematicity}c}).
Thus, the exact details of the electron bands in FeSe are very sensitive to small changes in the band structure parameters and there can be topologically significantly different Fermi surfaces arising from variations of the band positions of only a few meV in the presence of spin-orbit coupling as well as in high magnetic fields.

\subsection{Magnetotransport: small and tiny electron pockets}

At temperatures above 100~K, the magnetoresistance and Hall effect can be well-described as a compensated two carrier system \cite{Watson2015b}. The Hall effect is linear, and the hole and electron pockets have rather similar mobilities leading to a small overall value of the Hall coefficient which changes sign more than once as a function of temperature. However at temperatures below $\sim$80-90~K the Hall effect becomes noticeable non-linear, changing sign between the low and high magnetic field regimes ({\bf Figure~\ref{fig:QOs}c)}. This behavior can be described by going beyond the two-band model,
and considering the presence of an additional tiny electron band with  higher mobility than the either the hole band or the larger electron band \cite{Watson2015b,Huynh2016,Sun2016mob}.

The origin of a tiny electron pocket with a carrier density of about 0.7$\times 10^{20}$ cm$^{-3}$, a factor
5 smaller than the hole band, could originate as an inner electron band  {\bf Figure~\ref{fig:FS_evolution_nematicity}d} and give rise to the lowest frequencies in the quantum oscillations spectra ($\alpha_1$ and/or $\alpha_2$  in {\bf Figure~\ref{fig:QOs}b}).
Another scenario is that the presence of the small number of highly mobile carrier is linked to the {\it Dirac-like} dispersions in the nematic phase \cite{Tan2016} on some sections of the electron pockets \cite{Huynh2016}. As a result of the band shifts in the nematic phase, the $d_{xz}$ and $d_{xy}$ bands cross away from the M point near the ends of the peanut-shapes ({\bf Figure~\ref{fig:FS_evolution_nematicity}c}). However these band crossings are gapped due to spin-orbit coupling \cite{Watson2016}, (although this has been hard to resolve with ARPES), and near-linear dispersions exist only in a very limited regime. High mobility carriers have been detected across the whole nematic phase of FeSe$_{1-x}$S$_x$ \cite{Sun2016mob,Ovchenkov2017}.

However, the understanding of magnetotransport in FeSe is a subtle problem, and the complex large crossed-peanut shaped electron bands will have a strongly varying effective mass, and the Fermi surface curvature effects could play a role \cite{Ong1991}. Moreover, the scattering rate around the pocket could also vary strongly, being related to the orbital character on various sections. Anisotropic scattering rates were attributed to spin fluctuation scattering in Ref.~\cite{Tanatar2016}, and anisotropy in the relaxation time of excited electrons was detected in pump-probe measurements \cite{Luo2016_arxiv}. A large distribution of electron mobilities was also suggested by the mobility spectrum analysis in Ref.~\cite{Huynh2016}.

\subsection{Orbitally-dependent electronic correlations}

The Fe-chalcogenides are widely considered to be more strongly correlated than their Fe-pnictide cousins \cite{Yin2011,vanRoekeghem2016,Lanata2013,Yi2015_natcomm}. The effect of electronic correlations can be estimated by comparing
 DFT band structure calculations with the measured quasiparticle band dispersion in ARPES in the tetragonal phase of FeSe.
 It was found that the band renormalisation factors vary significantly
 between $\sim$ 3.2, 2.1 for the outer hole bands ($\alpha$,$\beta$) with $d_{xz/yz}$ orbital character,
 and a factor $\sim$8-9 for the $\gamma$ pocket, with $d_{xy}$ orbital character \cite{Maletz2014,Watson2015a}. However substantially larger band renormalisations have been observed in FeSe$_x$Te$_{1-x}$,  where  the band-selective renormalization
 can reach 17 \cite{Tamai2010,ZKLiu2015}, while correlations are expected to be weaker in FeS \cite{Miao2017}. The effective masses observed by quantum oscillations are also renormalised, with a particularly large mass enhancement on the outer electron pocket with largely $d_{xy}$ character \cite{Watson2015a,Coldea2016}. In electron-doped FeSe-based systems it has been shown that the $d_{xy}$ spectral weight depletes with increasing temperature due to particularly strong correlations effects \cite{Yi2015_natcomm}.

  The orbital selectivity of the observed renormalisations has been interpreted as evidence that the proximity to an orbital-selective Mott transition is a key to the understanding of the Fe-based superconductors \cite{Lanata2013}. In this picture, FeSe is a system in which the $d_{xy}$ orbital is closer to localisation than the other orbitals, due to its occupation being closer to half-filling.
  Calculations using DMFT on FeSe
give band renormalisations of $\sim$2.8 for the $d_{xz/yz}$ orbitals,
comparable to the measured values, although the
predicted value of $\sim$3.5 for the $d_{xy}$ band underestimates the experimental value \cite{Yin2011}. Moreover the incoherent {\it Hubbard bands} observed by ARPES at high binding energies are qualitatively captured by DMFT \cite{Watson2017a,Evtushinsky2016_arxiv}. However a narrowing of the bandwidth is also predicted from the inter-site Coulomb interaction $V$ \cite{Jiang2016}, and non-local correlations are likely to be important.

\section{FERMI SURFACE SHRINKING}

 Quantum oscillations, magnetotransport and ARPES agree on an important and unusual aspect of the electronic structure
of FeSe: the small size of the quasi-2D electron and hole pockets, which are much smaller than the expectations from DFT. In a compensated system, the total charge is conserved when the holes and electron pockets shrink simultaneously, but the origin of this effect is not yet established. Similar Fermi surface shrinking effects have been observed with quantum oscillations in e.g. LaFePO \cite{Coldea2008} and BaFe$_2$(As$_{1-x}$P$_x$)$_2$ \cite{Shishido2010}, which were sometimes quantified in terms of the typically 50-100 meV rigid shifts of the unrenrormalised DFT bands that would be required to account for the experimental Fermi surface area.

In FeSe, in order to bring the calculated Fermi surface close to the those determined experimentally, large shifts of up to
200~meV \cite{Watson2015a} would be needed. Momentum-dependent band shifts with opposite sign at the $\Gamma$ and M points have also been described from the ARPES perspective \cite{Borisenko2015}. A magnetic reconstruction of the Fermi surface can lead to small quantum oscillation frequencies, as seen in BaFe$_2$As$_2$ \cite{Terashima2011}. However in FeSe the nematic order gives in-plane distortions but does not dramatically affect the in-plane areas; the pockets are small even in the tetragonal phase \cite{Watson2015a,Watson2016}.

This shrinking effect has been suggested to be a natural consequence of the strong particle-hole asymmetry of
electronic bands, providing an indirect experimental evidence of strong interband scattering,
as a direct consequence of the coupling to a bosonic mode \cite{Ortenzi2009}. Recently, it has been suggested that in FeSe, due to the strong orbital-depending correlation effects, spin fluctuations between hole and electron pockets are responsible for an orbital-dependent shrinking of the Fermi surface that affects mainly the $xz/yz$ parts of the Fermi surface \cite{Fanfarillo2016}. An alternative perspective to explain the small and strongly renormalized low-energy band structure of FeSe is related to the consideration nearest neighbor Coulomb interaction, $V$. This interaction generates hopping corrections to the band dispersions and pushes the van Hove singularity at M point up towards the Fermi energy  while pulling down the top of the hole band at the $\Gamma$ point. The interatomic $V$ has also been discussed as a possible microscopic origin of the nematic ordering, where it is thought to favor a $d$-wave nematic bond order \cite{Jiang2016,Scherer2017}.

\subsection{Chemical potential effects in the tetragonal phase}

Recent temperature-dependent ARPES studies of FeSe up to room temperature report continuous temperature-dependent band shifts
up to $\sim $ 25~meV even in the tetragonal phase \cite{Rhodes2017_arxiv,AbdelHafiez2016,Kushnirenko2017_arxiv}.
Similar effects have been observed
other iron-based superconductors, such as Ba(Fe,Co)$_2$As$_2$ \cite{Brouet2013} and Ba(Fe,Ru)$_2$As$_2$ \cite{Dhaka2013},
being assigned to temperature-induced chemical potential shifts. In semimetals, where the top of the
hole bands and bottom of the electron bands are very close to
the chemical potential within the energy range of thermal broadening ($\mu \sim k_B T$), significant chemical potential shifts can occur due to balance the
thermal population of carriers. This effect
has interesting implications on the understanding of transport and magnetic measurements in this temperature range \cite{Rhodes2017_arxiv}.
The chemical potential shift may also have
influence at the nematic transition: a natural consequence of the nematic order parameter would be a shift of the chemical potential to
preserve charge, which might account for the $\sim$ 10 meV
momentum-independent downward shift of the $d_{xy}$ bands ({\bf Figure~\ref{fig:reviewfigarpestdepdata}}).

\section{TUNING NEMATIC ORDER WITH PHYSICAL AND CHEMICAL PRESSURE}

\begin{figure}
	\centering
	\includegraphics[width=0.8\linewidth]{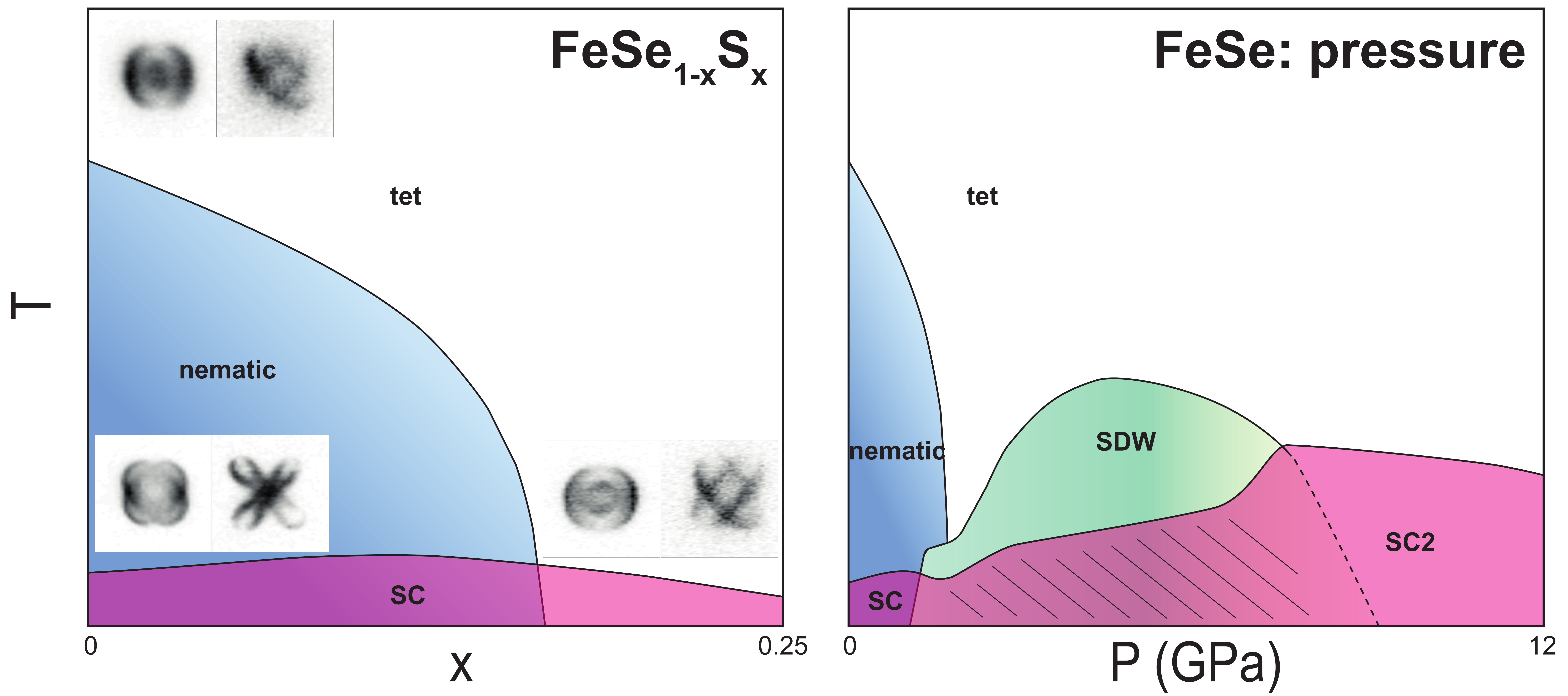}
	\caption{a) Schematic phase diagram of FeSe$_{1-x}$S$_x$, based on Refs.~\cite{Watson2015c,Coldea2016}. Nematicity is suppressed both as a function of temperature and sulphur substitution \cite{Watson2015c,Reiss2017}. b) Schematic phase diagram of FeSe under pressure, based on Ref.~\cite{Sun2015}. Hatched area could indicate a regime of coexisting superconductivity and antiferromagnetism \cite{Bendele2012}, while SC2 refers to the second dome of superconductivity with much higher $T_c$.}
	\label{fig:fesesx-schematicphasediagram}
\end{figure}

The nematic state of bulk FeSe can be tuned and suppressed either by
using applied hydrostatic pressures, isoelectronic substitution or chemical doping.
Applied hydrostatic pressure ({\bf Figure~\ref{fig:fesesx-schematicphasediagram}b}) initially suppresses nematicity  \cite{Medvedev2009,Terashima2015,Sun2016,Kothapalli2016}
and then stabilizes a new magnetic state \cite{Bendele2012,Kothapalli2016}, likely to be a stripe magnetic ordering. From this point onwards, the phase diagram resembles other Fe-based superconductors, with a high-$T_c$ phase being found once this magnetic order is suppressed at very high pressures \cite{Sun2016}.

Another tuning parameter of nematicity is isoelectronic substitution achieved by replacing selenium ions
for isoelectronic but smaller sulfur atoms in FeSe$_{1-x}$S$_x$, which acts as a chemical pressure effect \cite{Watson2015c,Mizuguchi2009}.
Surprisingly, this tuning parameter does not stabilize any long range magnetic order
outside the nematic phase  \cite{Watson2015c,Coldea2016}. Thus FeSe$_{1-x}$S$_x$ is an interesting and unique phase diagram in which the nematic transition temperature can be tuned to zero temperature without any magnetic order ({\bf Figure~\ref{fig:fesesx-schematicphasediagram}a)};
perhaps surprisingly the superconducting transition temperature change very little in response.

ARPES studies of FeSe$_{1-x}$S$_x$ showed a clear reduction in Fermi surface anisotropy and orbital ordering effects for both electron and hole Fermi surfaces \cite{Watson2015c}. For compositions beyond the nematic phase boundary, no anisotropies remain ({\bf Figure~\ref{fig:fesesx-schematicphasediagram}a}) \cite{Reiss2017}. Recent quantum oscillations studies of FeSe$_{1-x}$S$_x$ suggest a continuous expansion of the
outer electron and hole bands with chemical pressure even outside the nematic state \cite{Coldea2016}. This is in contrast to the
observation of only small Fermi surfaces in FeSe under pressure, suggested to result from magnetic reconstruction \cite{Terashima2015}.
Nematic susceptibility studies as a function of chemical pressure
have revealed the possible presence of a nematic critical
point \cite{Hosoi2016}. However further evidence for quantum criticality, such as a divergent quasiparticle effective masses in quantum oscillations \cite{Coldea2016} or thermodynamic probes \cite{Abdel-Hafiez2015}, remains to be found.

Another interesting aspect of the electronic structure of FeSe$_{1-x}$S$_x$ are the possible changes of the Fermi surface topology, resulting from the suppression of nematic ordering. The inner hole pocket which is pushed below the chemical potential by nematic ordering in FeSe emerges at low temperatures in ARPES data for substitutions higher than 11\% \cite{Watson2015c}, while the electron pockets lose their in-plane distortions. Quantum oscillations measurements in very high magnetic fields detect a prominent low frequency oscillation around $x=0.12$ \cite{Coldea2016} that could tentatively be linked with the re-emerging inner hole pocket ({\bf Figure~\ref{fig:FS_evolution_nematicity}e}). However, this large amplitude low frequency oscillation is not detected at higher sulphur substitution beyond the nematic phase \cite{Coldea2016}. This implies that there could be other topological changes in the Fermi surface, a possible Liftshitz transition,
 as a function of chemical pressure, involving other small electron bands or breakdown orbits  \cite{Coldea2016}. Hall effect measurements in low fields suggest that
 the high mobility electrons  seem to disappear outside the nematic state in  FeSe$_{1-x}$S$_x$ \cite{Ovchenkov2016}.

\section{THE INTERPLAY OF NEMATIC AND SUPERCONDUCTING ORDERS}

The normal state nematic electronic structure of FeSe with highly anisotropic Fermi surfaces has profound implications on the superconducting pairing and the symmetry of the superconducting order parameter. A highly twofold anisotropic superconducting gap has been recently found by ARPES in FeSe$_{0.93}$S$_{0.07}$ \cite{Xu2016}.
Moreover recent Bogoliubov quasiparticle scattering
interference imaging at very low temperatures also
found an extremely anisotropic superconducting gap in FeSe. The gap structure is
nodeless, though the gap reaches rather small values on the major axis of the elliptical hole pocket, and the gap has opposite sign between the hole and
the small electron pockets \cite{Sprau2016_arxiv}. Although there has been some reports suggesting the presence of nodes in the superconducting gap of FeSe \cite{Song2011,Moore2015,Kasahara2014}, most of the thermodynamic and thermal conductivity studies of bulk FeSe in the superconducting phase can be explained
if at least two different nodeless superconducting gaps are present
\cite{Lin2011,Bourgeois-Hope2016}.

The strong anisotropy of the superconducting gap is an important indication of unconventional pairing in FeSe. It has been suggested that the anisotropic gap is due to an orbital-dependent pairing mechanism, with the maximum gap on sections with $d_{yz}$ character, and a small gap on sections with $d_{xz}$ or $d_{xy}$ character \cite{Kreisel2016_arxiv,Sprau2016_arxiv}. In this picture, superconductivity results from the nesting of $d_{yz}$ sections of the hole and electron bands which couples strongly to ($\pi,0$) magnetic fluctuations, whereas the $d_{xy}$ character does not participate.

Due to the presence of a small electron band with small Fermi energy
and comparable to the superconducting gap is suggested
that FeSe could be placed into BCS-BEC cross-over regime \cite{Kasahara2014}.
Furthermore, as the Zeeman energy for this small electron band also becomes comparable
to its Fermi energy and gap energy, a highly spin-polarized, pairing superconductivity is
inferred to form in high magnetic fields
at low temperatures  \cite{Watashige2017}.

As a function of chemical pressure in FeSe$_{1-x}$S$_x$, the presence of multi-gap superconductivity is preserved
\cite{Abdel-Hafiez2015} whereas tunnelling experiments
found that the vortex core anisotropy is strongly suppressed \cite{Moore2015}.
In FeSe, high-resolution thermal expansion
 showed a lack of coupling between the orthorhombicity and superconductivity \cite{Bohmer2013}. However in samples with up to 15\% sulphur substitution and correspondingly reduced tetragonal-to-orthorhombic transition temperatures, an enhancement of the orthorhombic distortion
 in the superconducting state \cite{Wang2016Meingast} was observed. This may indeed indicate that superconductivity favors the nematic state.

The interplay of nematic, superconducting and magnetic orders has attracted much interest. There is a large body of recent theoretical work on FeSe addressing the competing instabilities of the nematic order
in relation to spin-density wave and superconductivity, as detailed in Refs. \cite{Xing2017,Kreisel2015,Yamakawa2016},
as well as on the role played by strong inter-site Coulomb interactions  \cite{Jiang2016,Scherer2017}.

\section{CONCLUSION}

FeSe has opened up new avenues towards understanding unconventional superconductivity in Fe-based superconductors. Experimental efforts on this simple system have started to reveal the key ingredients responsible for its still mysterious nematic behavior and its the highly tunable superconductivity, while exposing its complexities. Insights have been gained from various experimental probes, which are now pointing towards a common picture: FeSe is a system where strong correlations, orbital-selectivity, spin-orbit coupling, nematic bond ordering and fluctuating magnetism are all important. FeSe is perhaps the cleanest example of a nematic phase in condensed matter, with the small size of the Fermi surface pockets and small Fermi energies of the bands being distinguishing features of its multi-band, multi-orbital electronic structure.

\begin{summary}[SUMMARY POINTS]
	\begin{enumerate}
		\item FeSe is a fascinating quantum material which displays a
unique nematic electronic state, from which a highly anisotropic superconductivity emerges.
		\item The nematic state of FeSe is characterised by the development of an unusual momentum-dependent orbital ordering, described by a bond ordering and not on-site occupations.
		\item  FeSe is a strongly correlated system, belonging to the class of iron chalcogenides, in which
orbital-dependent quasiparticle mass renormalizations are important, with the $d_{xy}$ orbitals being the most correlated.
		\item The Fermi surface of FeSe at low temperature is that of
a compensated semimetal, with one small elliptical hole band and two electron bands, one of the electron bands having
an unusually high mobility.
\item The sizes of the quasi-2D Fermi surfaces are much smaller than predicted by \textit{ab-initio} calculations, likely linked to non-local interactions such as the inter-site Coulomb repulsion.
\item The low energy electronic structure of FeSe
reveals many small comparable energy scales: spin-orbit coupling, nematic order, the effective Fermi energies of bands, and superconductivity.
\item Tuning the system with chemical and physical pressure can lead to quite dramatic effects: induced magnetism and high-$T_c$ are found under pressure, while the suppression of nematic order does not lead to any strongly enhanced $T_c$ in FeSe$_{1-x}$S$_x$.
	\end{enumerate}
\end{summary}

\begin{issues}[FUTURE ISSUES]
	\begin{enumerate}
		\item Is the origin of the anisotropy of the superconducting gap in FeSe
a simple consequence of its nematic electronic structure  or is
a manifestation of a highly anisotropic orbitally-dependent pairing mechanism ?
		\item Are nematic fluctuations important for enhancing superconductivity
in FeSe under pressure or the spin fluctuations are always in the driving seat ?
\item What is the role played by the small electronic bands with high mobility for superconductivity in FeSe ?
		\item Can chemical and applied pressure in FeSe help to disentangle
the role of nematicity, magnetism, orbital order and electronic correlations in FeSe
and identify the direct route towards high-$T_c$ superconductivity ?		
		\item Can high-resolution detwinned ARPES measurements of FeSe finally settle the open questions regarding the orbital order parameter in FeSe?
				\item The large 50 meV energy scale seen in ARPES in FeSe is very similar to previous results in NaFeAs and BaFe$_2$As$_2$. Is this a universal feature of the nematic phase, independent of magnetic order?
			\end{enumerate}
\end{issues}

\section*{DISCLOSURE STATEMENT}
The authors are not aware of any affiliations, memberships, funding, or financial holdings that
might be perceived as affecting the objectivity of this review.

\section*{ACKNOWLEDGMENTS}
We are very grateful to all our numerous collaborators for their important scientific contribution to the understanding
of the electronic structure of FeSe. 
Special thanks go to Timur Kim and Moritz Hoesch,
for the development of I05 and technical support which was instrumental in obtaining the high-resolution ARPES data on FeSe.
We are also very thankful to  Amir Haghighirad, Thomas Wolf and Shigeru Kasahara for the growth of high quality crystals,
Andy Schofield, Roser Valenti, Steffen Backes for theoretical support, 
Andrey Chubukov, Peter Hirschfeld, Brian Anderson, Rafael Fernandes and Oskar Vafek, A. Kreisel, Ilya Eremin and Takasada Shibauchi for numerous insightful discussions. We thank Alix McCollam, David Graf, Eung San, William Knafo, Pascal Reiss, Samuel Blake, Mara Bruma for their contributions to high magnetic field studies as well as Puning Zhao for technical support. We acknowledge the financial support provided by EPSRC  (EP/I004475/1,  EP/I017836/1)
and Diamond  Light  Source  for  experimental access  to  the I05 Beamline.
Amalia Coldea is grateful for an EPSRC Career Acceleration Fellowship (EP/I004475/1).

%
\section*{LITERATURE\ CITED}
\bibliographystyle{ar-style4}

\begin{thebibliography}{104}
\expandafter\ifx\csname natexlab\endcsname\relax\def\natexlab#1{#1}\fi

\bibitem{Kamihara2008}
Kamihara Y, Watanabe T, Hirano M, Hosono H. 2008.
\textit{J. Am. Chem. Soc.} 130:3296

\bibitem{Hsu2008}
Hsu FC, Luo JY, Yeh KW, Chen TK, Huang TW, et~al. 2008.
\textit{Proceedings of the National Academy of Sciences} 105:14262--14264

\bibitem{Medvedev2009}
Medvedev S, McQueen TM, Troyan IA, Palasyuk T, Eremets MI, et~al. 2009.
\textit{Nat. Mater.} 8:630

\bibitem{Lei2016}
Lei B, Cui J, Xiang Z, Shang C, Wang N, et~al. 2016.
\textit{Physical Review Letters} 116:077002

\bibitem{Sun2015}
Sun H, Woodruff DN, Cassidy SJ, Allcroft GM, Sedlmaier SJ, et~al. 2015.
\textit{Inorganic Chemistry} 54:1958--1964

\bibitem{BurrardLucas2013}
Burrard-Lucas M, Free DG, Sedlmaier SJ, Wright JD, Cassidy SJ, et~al. 2013.
\textit{Nature Materials} 12:15

\bibitem{Wang2012monolayer}
Qing-Yan W, Zhi L, Wen-Hao Z, Zuo-Cheng Z, Jin-Song Z, et~al. 2012.
\textit{Chinese Physics Letters} 29:037402

\bibitem{Wang2017review}
Wang Z, Liu C, Liu Y, Wang J. 2017.
\textit{Journal of Physics: Condensed Matter} 29:153001

\bibitem{Bohmer2013}
B{\"{o}}hmer AE, Hardy F, Eilers F, Ernst D, Adelmann P, et~al. 2013.
\textit{Phys. Rev. B} 87:180505

\bibitem{Chareev2013}
Chareev D, Osadchii E, Kuzmicheva T, Lin JY, Kuzmichev S, et~al. 2013.
\textit{CrystEngComm} 15:1989

\bibitem{Bohmer2016}
B\"ohmer AE, Taufour V, Straszheim WE, Wolf T, Canfield PC. 2016.
\textit{Phys. Rev. B} 94:024526

\bibitem{Watson2015a}
Watson MD, Kim TK, Haghighirad AA, Davies NR, McCollam A, et~al.
  2015{\natexlab{a}}.
\textit{Phys. Rev. B} 91:155106

\bibitem{Watson2016}
Watson MD, Kim TK, Rhodes LC, Eschrig M, Hoesch M, et~al. 2016.
\textit{Phys. Rev. B} 94:201107

\bibitem{Rhodes2017_arxiv}
{Rhodes} LC, {Watson} MD, {Haghighirad} AA, {Eschrig} M, {Kim} TK. 2017.
\textit{arXiv} :1702.06321

\bibitem{Gunnarsson2003}
Gunnarsson O, Calandra M, Han JE. 2003.
\textit{Rev. Mod. Phys.} 75:1085

\bibitem{Mcqueen2009a}
McQueen TM, Williams AJ, Stephens PW, Tao J, Zhu Y, et~al. 2009.
\textit{Phys. Rev. Lett.} 103:057002

\bibitem{Khasanov2010}
Khasanov R, Bendele M, Conder K, Keller H, Pomjakushina E, Pomjakushin V. 2010.
\textit{New J. Phys.} 12:073024

\bibitem{Chubukov2015}
Chubukov AV, Fernandes RM, Schmalian J. 2015.
\textit{Phys. Rev. B} 91:201105

\bibitem{Fernandes2014}
Fernandes RM, Chubukov AV, Schmalian J. 2014.
\textit{Nat. Phys.} 10:97

\bibitem{Onari2016}
Onari S, Yamakawa Y, Kontani H. 2016.
\textit{Phys. Rev. Lett.} 116:227001

\bibitem{Wang2015}
Wang F, Kivelson SA, Lee DH. 2015.
\textit{Nature Physics} 11:959

\bibitem{Tanatar2016}
Tanatar MA, B\"ohmer AE, Timmons EI, Sch\"utt M, Drachuck G, et~al. 2016.
\textit{Phys. Rev. Lett.} 117:127001

\bibitem{Kasahara2014}
Kasahara S, Watashige T, Hanaguri T, Kohsaka Y, Yamashita T, et~al. 2014.
\textit{Proc. Natl. Acad. Sci. U. S. A.} 111:16309

\bibitem{Sprau2016_arxiv}
{Sprau} PO, {Kostin} A, {Kreisel} A, {B{\"o}hmer} AE, {Taufour} V, et~al. 2016.
\textit{arXiv} :1611.02134

\bibitem{Hosoi2016}
Hosoi S, Matsuura K, Ishida K, Wang H, Mizukami Y, et~al. 2016.
\textit{PNAS} 113:8139

\bibitem{Pomeranchuk1959}
Pomeranchuk II. 1959.
\textit{JETP} 8:361

\bibitem{Massat2016}
Massat P, Farina D, Paul I, Karlsson S, Strobel P, et~al. 2016.
\textit{Proc. Natl. Acad. Sci.} 113:9177

\bibitem{Wang2016}
Wang Q, Shen Y, Pan B, Hao Y, Ma M, et~al. 2016{\natexlab{a}}.
\textit{Nat Mater} 15:159

\bibitem{Glasbrenner2015}
Glasbrenner JK, Mazin II, Jeschke HO, Hirschfeld P, Fernandes RM, Valent{\'\i}
  R. 2015.
\textit{Nature Physics} 11:953

\bibitem{Chubukov2016}
Chubukov AV, Khodas M, Fernandes RM. 2016.
\textit{Phys. Rev. X} 6:041045

\bibitem{Xu2016}
Xu HC, Niu XH, Xu DF, Jiang J, Yao Q, et~al. 2016.
\textit{Phys. Rev. Lett.} 117:157003

\bibitem{YWang2015}
Wang Y, Berlijn T, Hirschfeld PJ, Scalapino DJ, Maier TA. 2015.
\textit{Phys. Rev. Lett.} 114:107002

\bibitem{Tomic2014}
Tomi{\'{c}} M, Jeschke HO, Valent{\'{i}} R. 2014.
\textit{Phys. Rev. B} 90:195121

\bibitem{Fedorov2016}
Fedorov A, Yaresko A, Kim TK, Kushnirenko Y, Haubold E, et~al. 2016.
\textit{Sci. Rep.} 6:36834

\bibitem{Borisenko2015}
Borisenko SV, Evtushinsky DV, Liu ZH, Morozov I, Kappenberger R, et~al. 2016.
\textit{Nat Phys} 12:311--317

\bibitem{Brouet2012}
Brouet V, Jensen MF, Lin PH, Taleb-Ibrahimi A, {Le F{\`{e}}vre} P, et~al. 2012.
\textit{Phys. Rev. B} 86:075123

\bibitem{Moreschini2014}
Moreschini L, Lin PH, Lin CH, Ku W, Innocenti D, et~al. 2014.
\textit{Phys. Rev. Lett.} 112:087602

\bibitem{Pustovit2016}
Pustovit YV, Kordyuk AA. 2016.
\textit{Low Temperature Physics} 42:995

\bibitem{Liu2015a}
Liu X, Zhao L, He S, He J, Liu D, et~al. 2015{\natexlab{a}}.
\textit{J. Phys. Condens. Matter} 27:183201

\bibitem{Huang2017_arxiv}
{Huang} D, {Hoffman} JE. 2017.
\textit{arXiv} :1703.09306

\bibitem{Richard2011}
Richard P, Sato T, Nakayama K, Takahashi T, Ding H. 2011.
\textit{Reports on Progress in Physics} 74:124512

\bibitem{Kordyuk2012}
Kordyuk AA. 2012.
\textit{Low Temperature Physics} 38:888

\bibitem{Richard2015}
Richard P, Qian T, Ding H. 2015.
\textit{Journal of Physics: Condensed Matter} 27:293203

\bibitem{vanRoekeghem2016}
{van Roekeghem} A, Richard P, Ding H, Biermann S. 2016.
\textit{Comptes Rendus Physique} 17:140

\bibitem{Maletz2014}
Maletz J, Zabolotnyy VB, Evtushinsky DV, Thirupathaiah S, Wolter AUB, et~al.
  2014.
\textit{Phys. Rev. B} 89:220506

\bibitem{Fernandes2014b}
Fernandes RM, Vafek O. 2014.
\textit{Phys. Rev. B} 90:214514

\bibitem{Watson2017c_arxiv}
{Watson} MD, {Haghighirad} AA, {Takita} H, {Mansur} W, {Iwasawa} H, et~al.
  2017.
\textit{arXiv} :1702.05460

\bibitem{Shimojima2014}
Shimojima T, Suzuki Y, Sonobe T, Nakamura A, Sakano M, et~al. 2014.
\textit{Phys. Rev. B} 90:121111

\bibitem{Suzuki2015}
Suzuki Y, Shimojima T, Sonobe T, Nakamura A, Sakano M, et~al. 2015.
\textit{Phys. Rev. B} 92:205117

\bibitem{Tan2013}
Tan S, Zhang Y, Xia M, Ye Z, Chen F, et~al. 2013.
\textit{Nat. Mater.} 12:634--40

\bibitem{Nakayama2014}
Nakayama K, Miyata Y, Phan GN, Sato T, Tanabe Y, et~al. 2014.
\textit{Phys. Rev. Lett.} 113:237001

\bibitem{Fanfarillo2016}
Fanfarillo L, Mansart J, Toulemonde P, Cercellier H, Le~F\`evre P, et~al. 2016.
\textit{Phys. Rev. B} 94:155138

\bibitem{Zhang2016}
Zhang Y, Yi M, Liu ZK, Li W, Lee JJ, et~al. 2016.
\textit{Phys. Rev. B} 94:115153

\bibitem{Watson2015b}
Watson MD, Yamashita T, Kasahara S, Knafo W, Nardone M, et~al.
  2015{\natexlab{b}}.
\textit{Phys. Rev. Lett.} 115:027006

\bibitem{Kreisel2015}
Kreisel A, Mukherjee S, Hirschfeld PJ, Andersen BM. 2015.
\textit{Phys. Rev. B} 92:224515

\bibitem{Jiang2016}
Jiang K, Hu J, Ding H, Wang Z. 2016.
\textit{Phys. Rev. B} 93:115138

\bibitem{Yi2015}
Yi L, Xian-Xin W, Jiang-Ping H. 2015.
\textit{Chinese Physics Letters} 32:117402

\bibitem{Zhang2015}
Zhang P, Qian T, Richard P, Wang XP, Miao H, et~al. 2015.
\textit{Phys. Rev. B} 91:214503

\bibitem{Scherer2017}
Scherer DD, Jacko AC, Friedrich C, \ifmmode \mbox{\c{S}}\else
  \c{S}\fi{}a\ifmmode \mbox{\c{s}}\else \c{s}\fi{}\ifmmode \imath \else \i
  \fi{}o\ifmmode~\breve{g}\else \u{g}\fi{}lu E, Bl\"ugel S, et~al. 2017.
\textit{Phys. Rev. B} 95:094504

\bibitem{Bohmer2014}
B{\"{o}}hmer AE, Arai T, Hardy F, Hattori T, Iye T, et~al. 2015.
\textit{Phys. Rev. Lett.} 114:027001

\bibitem{Watson2017a}
Watson MD, Backes S, Haghighirad AA, Hoesch M, Kim TK, et~al. 2017.
\textit{Phys. Rev. B} 95:081106

\bibitem{Evtushinsky2016_arxiv}
{Evtushinsky} DV, {Aichhorn} M, {Sassa} Y, {Liu} ZH, {Maletz} J, et~al. 2016.
\textit{ArXiv e-prints}

\bibitem{Terashima2014}
Terashima T, Kikugawa N, Kiswandhi A, Choi ES, Brooks JS, et~al. 2014.
\textit{Phys. Rev. B} 90:144517

\bibitem{Audouard2014}
Audouard A, Duc F, Drigo L, Toulemonde P, Karlsson S, et~al. 2015.
\textit{Europhys. Lett.} 109:27003

\bibitem{Coldea2016}
Coldea AI, Blake SF, Kasahara S, Haghighirad AA, Watson MD, et~al. 2017.
\textit{arXiv:1611.07424}

\bibitem{Reiss2017}
Reiss P, Watson MD, Kim TK, Haghighirad A. A.~Woodruff DN, Bruma M, et~al.
  2017.
\textit{arXiv:1705.11139}

\bibitem{Coldea2008}
Coldea AI, Fletcher JD, Carrington A, Analytis JG, Bangura AF, et~al. 2008.
\textit{Phys. Rev. Lett.} 101:216402

\bibitem{Putzke2012}
Putzke C, Coldea AI, Guillam\'on I, Vignolles D, McCollam A, et~al. 2012.
\textit{Phys. Rev. Lett.} 108:047002

\bibitem{Huynh2016}
Huynh KK, Tanabe Y, Urata T, Oguro H, Heguri S, et~al. 2014.
\textit{Phys. Rev. B} 90:144516

\bibitem{Sun2016mob}
Sun Y, Pyon S, Tamegai T. 2016{\natexlab{a}}.
\textit{Physical Review B} 93:104502

\bibitem{Tan2016}
Tan SY, Fang Y, Xie DH, Feng W, Wen CHP, et~al. 2016.
\textit{Phys. Rev. B} 93:104513

\bibitem{Ovchenkov2017}
Ovchenkov YA, Chareev DA, Kulbachinskii VA, Kytin VG, Presnov DE, et~al. 2017.
\textit{Superconductor Science and Technology} 30:035017

\bibitem{Ong1991}
Ong NP. 1991.
\textit{Phys. Rev. B} 43:193

\bibitem{Luo2016_arxiv}
{Luo} CW, {Cheng} PC, {Wang} SH, {Chiang} JC, {Lin} J, et~al. 2016.
\textit{arXiv} :1603.08710

\bibitem{Yin2011}
Yin ZP, Haule K, Kotliar G. 2011.
\textit{Nat. Mater.} 10:932--5

\bibitem{Lanata2013}
Lanat\`a N, Strand HUR, Giovannetti G, Hellsing B, de' Medici L, Capone M.
  2013.
\textit{Phys. Rev. B} 87:045122

\bibitem{Yi2015_natcomm}
Yi M, Liu Z, Zhang Y, Yu R, Zhu J, et~al. 2015.
\textit{Nat. Comm.} 6:7777

\bibitem{Tamai2010}
Tamai A, Ganin AY, Rozbicki E, Bacsa J, Meevasana W, et~al. 2010.
\textit{Phys. Rev. Lett.} 104:097002

\bibitem{ZKLiu2015}
Liu ZK, Yi M, Zhang Y, Hu J, Yu R, et~al. 2015{\natexlab{b}}.
\textit{Phys. Rev. B} 92:235138

\bibitem{Miao2017}
Miao J, Niu XH, Xu DF, Yao Q, Chen QY, et~al. 2017.
\textit{arXiv} :1703.08682

\bibitem{Shishido2010}
Shishido H, Bangura AF, Coldea AI, Tonegawa S, Hashimoto K, et~al. 2010.
\textit{Phys. Rev. Lett.} 104:1

\bibitem{Terashima2011}
Terashima T, Kurita N, Tomita M, Kihou K, Lee CH, et~al. 2011.
\textit{Phys. Rev. Lett.} 107:176402

\bibitem{Ortenzi2009}
Ortenzi L, Cappelluti E, Benfatto L, Pietronero L. 2009.
\textit{Phys. Rev. Lett.} 103:046404

\bibitem{AbdelHafiez2016}
Abdel-Hafiez M, Pu YJ, Brisbois J, Peng R, Feng DL, et~al. 2016.
\textit{Phys. Rev. B} 93:224508

\bibitem{Kushnirenko2017_arxiv}
{Kushnirenko} Y, {Kordyuk} AA, {Fedorov} A, {Haubold} E, {Wolf} T, et~al. 2017.
\textit{arXiv} :1702.02088

\bibitem{Brouet2013}
Brouet V, Lin PH, Texier Y, Bobroff J, Taleb-Ibrahimi A, et~al. 2013.
\textit{Phys. Rev. Lett.} 110:167002

\bibitem{Dhaka2013}
Dhaka RS, Hahn SE, Razzoli E, Jiang R, Shi M, et~al. 2013.
\textit{Phys. Rev. Lett.} 110:067002

\bibitem{Watson2015c}
Watson MD, Kim TK, Haghighirad AA, Blake SF, Davies NR, et~al.
  2015{\natexlab{c}}.
\textit{Phys. Rev. B} 92:121108

\bibitem{Bendele2012}
Bendele M, Ichsanow A, Pashkevich Y, Keller L, Str\"assle T, et~al. 2012.
\textit{Phys. Rev. B} 85:064517

\bibitem{Terashima2015}
Terashima T, Kikugawa N, Kasahara S, Watashige T, Shibauchi T, et~al. 2015.
\textit{J. Phys. Soc. Japan} 84:063701

\bibitem{Sun2016}
Sun Y, Pyon S, Tamegai T. 2016{\natexlab{b}}.
\textit{Physical Review B} 93:104502

\bibitem{Kothapalli2016}
Kothapalli K, B{\"{o}}hmer AE, Jayasekara WT, Ueland BG, Das P, et~al. 2016.
\textit{Nature Communications} 7:12728

\bibitem{Mizuguchi2009}
Mizuguchi Y, Tomioka F, Tsuda S, Yamaguchi T, Takano Y. 2009.
\textit{Journal of the Physical Society of Japan} 78:074712

\bibitem{Abdel-Hafiez2015}
Abdel-Hafiez M, Zhang YY, Cao ZY, Duan CG, Karapetrov G, et~al. 2015.
\textit{Phys. Rev. B} 91:165109

\bibitem{Ovchenkov2016}
Ovchenkov Y, Chareev D, Kulbachinskii VA, Kytin V, Presnov D, et~al. 2016.
\textit{arXiv:1607.05669}

\bibitem{Song2011}
Song CL, Wang YL, Cheng P, Jiang YP, Li W, et~al. 2011.
\textit{Science} 332:1410

\bibitem{Moore2015}
Moore SA, Curtis JL, Di~Giorgio C, Lechner E, Abdel-Hafiez M, et~al. 2015.
\textit{Phys. Rev. B} 92:235113

\bibitem{Lin2011}
Lin JY, Hsieh YS, Chareev DA, Vasiliev AN, Parsons Y, Yang HD. 2011.
\textit{Phys. Rev. B} 84:220507

\bibitem{Bourgeois-Hope2016}
Bourgeois-Hope P, Chi S, Bonn DA, Liang R, Hardy WN, et~al. 2016.
\textit{Phys. Rev. Lett.} 117:097003

\bibitem{Kreisel2016_arxiv}
{Kreisel} A, {Andersen} BM, {Sprau} PO, {Kostin} A, {S{\'e}amus Davis} JC,
  {Hirschfeld} PJ. 2016.
\textit{arXiv 1611.02643}

\bibitem{Watashige2017}
Watashige T, Arsenijevi{\'{c}} S, Yamashita T, Terazawa D, Onishi T, et~al.
  2017.
\textit{Journal of the Physical Society of Japan} 86:014707

\bibitem{Wang2016Meingast}
Wang L, Hardy F, Wolf T, Adelmann P, Fromknecht R, et~al. 2016{\natexlab{b}}.
\textit{physica status solidi (b)} 1-6

\bibitem{Xing2017}
Xing RQ, Classen L, Khodas M, Chubukov AV. 2017.
\textit{Phys. Rev. B} 95:085108

\bibitem{Yamakawa2016}
Yamakawa Y, Onari S, Kontani H. 2016.
\textit{Phys. Rev. X} 6:021032

\end{thebibliography}

\end{document}